\pdfoutput=1
\documentclass[acmtog]{acmart}

\makeatletter                   
\def\mdseries@tt{m}             
\makeatother                    

\acmSubmissionID{529}
\usepackage{graphicx}
\graphicspath{{figure}, {images}, {example}}
\DeclareGraphicsExtensions{.png,.PNG,.jpeg,.JPEG,.jpg,.JPG,.bmp,.BMP}
\usepackage{epsfig} 
\usepackage{subcaption}
\usepackage[export]{adjustbox}
\usepackage{float}
\usepackage[justification=raggedright]{caption}	
\usepackage{lscape} 
\usepackage{overpic}

\usepackage[lined,ruled,linesnumbered]{algorithm2e}

\SetAlFnt{\small}
\SetAlCapFnt{\small}
\SetAlCapNameFnt{\small}
\SetAlCapHSkip{0pt}
\IncMargin{-\parindent}

\usepackage{booktabs}  
\usepackage{multirow}
\usepackage{paralist}
\usepackage{enumitem}

\usepackage{mathtools}
\usepackage{amssymb,amsmath}   

\usepackage{units}
\usepackage{color}

\usepackage{comment}

\def\etal{et al.}			  
\def\eg{e.g.,~}               
\def\ie{i.e.,~}               

\newlength\paramargin
\newlength\figmargin
\newlength\secmargin
\newlength\figcapmargin

\newcommand{\mpage}[2]
{
\begin{minipage}{#1\linewidth}\centering
#2
\end{minipage}
}

\newcommand{\secref}[1]{Section~\ref{sec:#1}}
\newcommand{\figref}[1]{Figure~\ref{fig:#1}} 
\newcommand{\tblref}[1]{Table~\ref{tbl:#1}}
\newcommand{\eqnref}[1]{Equation~\ref{eq:#1}}

\long\def\ignorethis#1{}
\newcommand{\james}[1]{\textcolor{blue}{[\textbf{James}: #1]}} 
\newcommand{\LJ}[1]{{\textcolor{olive}{[\textbf{LJ:} #1]}}}

\newcommand{\tb}[1]{\textbf{#1}}

\newcommand{\hl}[1]{\textcolor{red}{#1}}

\def\imgSymbol{I}
\def\inputI{\mathbf{\imgSymbol}}            
\newcommand{\inputIk}[1]{\imgSymbol_{#1}}   

\def\segSymbol{M}
\def\segI{\mathbf{\segSymbol}}                  
\newcommand{\segIk}[1]{\segSymbol_{#1}}         

\def\rotSymbol{R}
\newcommand{\camRk}[1]{\rotSymbol_{#1}}          

\def\tranSymbol{T}
\newcommand{\camTk}[1]{\tranSymbol_{#1}}        

\def\skelSymbol{c}
\def\skelC{\mathbf{\skelSymbol}}                
\newcommand{\skelCk}[1]{\skelSymbol_{#1}}       

\def\frameNum{K}                                
\def\curvePntNum{m}                             
\def\skelPntNum{n}                              

\def\curveSymbol{C}
\def\curveStruct{\curveSymbol}
\newcommand{\curveProjk}[1]{c'_{#1}}

\def\occluFunc{\mathbb{I}_{occ}}

\def\RE{RE} 
\def\RRE{RRE} 
\def\RPE{RPE} 
\def\PE{PE} 
\def\TPE{TPE} 
\def\TRE{TRE} 

\setcopyright{acmcopyright}
\acmJournal{TOG}
\acmYear{2020}
\acmVolume{00}
\acmNumber{0}
\acmArticle{0}
\acmMonth{0}

\acmDOI{0000001.0000001_2}

\begin{document}
\title{Vid2Curve: Simultaneous Camera Motion Estimation and Thin Structure Reconstruction from an RGB Video}

\author{Peng Wang}
\affiliation{%
  \institution{The University of Hong Kong}
} 
\author{Lingjie Liu}
\affiliation{%
  \institution{Max Planck Institute for Informatics}
}
\author{Nenglun Chen}
\affiliation{%
  \institution{The University of Hong Kong}
}
\author{Hung-Kuo Chu}
\affiliation{%
  \institution{National Tsing Hua University}
}
\author{Christian Theobalt}
\affiliation{%
  \institution{Max Planck Institute for Informatics}
}
\author{Wenping Wang}
\affiliation{%
  \institution{The University of Hong Kong}
}


\begin{abstract}
 Thin structures, such as wire-frame sculptures, fences, cables, power lines, and tree branches, are common in the real world. 
 %
 It is extremely challenging to acquire their 3D digital models using traditional image-based or depth-based reconstruction methods, because thin structures often lack distinct point features and have severe self-occlusion.  
 We propose the first approach that simultaneously estimates camera motion and reconstructs the geometry of complex 3D thin structures in high quality from a color video captured by a handheld camera. 
 %
 Specifically, we present a new curve-based approach to estimate accurate camera poses by establishing correspondences between featureless thin objects in the foreground in consecutive video frames, without requiring visual texture in the background scene to lock on. 
 %
 Enabled by this effective curve-based camera pose estimation strategy, we develop an iterative optimization method with tailored measures on geometry, topology as well as self-occlusion handling for reconstructing 3D thin structures.
 %
 Extensive validations on a variety of thin structures show that our method achieves accurate camera pose estimation and faithful reconstruction of 3D thin structures with complex shape and topology at a level that has not been attained by other existing reconstruction methods. 
\end{abstract}

%
%


%
%


\begin{teaserfigure}
  \includegraphics[width=\linewidth]{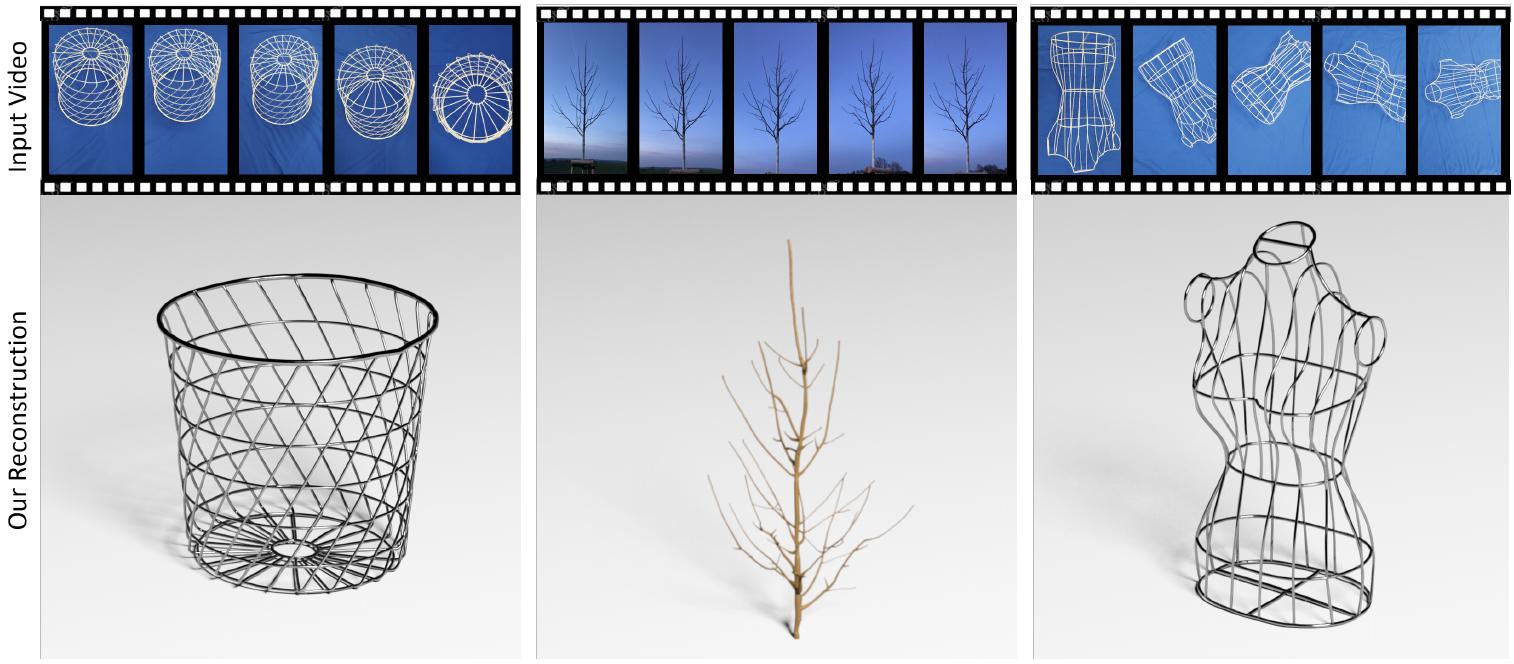}
  \caption{This figure shows three thin structured objects reconstructed using our method, together with selected frames of the input videos. Using an RGB video as input, our method performs curve-based camera pose estimation and reconstructs complex 3D thin structures in better quality than other existing methods. See the comparisons with the other methods on the bucket model and the hanger model in Section \ref{sec:baseline_compare}. 
%
  }
  \label{fig:teaser}
\end{teaserfigure}

\maketitle

\section{Introduction}
\label{sec:intro}
The real world is full of objects made of thin elements, such as wire-frame sculptures, fences, power lines, leafless trees, etc. Reconstructing 3D geometry of these thin objects is a critical task in various applications. For example, wire art is widely used in the design of furniture, jewelry, and sculptures.
Accurate 3D scanning of wire art is important for cultural heritage preservation and for filling virtual worlds with authentic contents.   
Reconstruction of (leafless) trees helps robot-assisted fruit tree pruning in agriculture, and power line reconstruction helps aerial drones avoid collisions. 

It is well known that reconstructing thin structures is very difficult with either image-based or depth-based methods. Traditional image-based methods perform 3D reconstruction by estimating camera poses using visual feature points in the scene and then identifying the dense point correspondences between texture features. Since thin structures lack texture features and are only a few pixels wide, classical correspondence matching methods
perform poorly or fail. Besides, even small camera calibration errors may severely compromise the reconstruction accuracy of thin structures. 
To address this issue, some recent works leverage high-order curve primitives for image matching~\cite{Xiao:josaa:05,Fabbri:cvpr:2010,Nurutdinova:iccv:2015,Usumezbas:eccv:16}. However, these works assume input images are pre-calibrated. This assumption is hard to meet for thin structure scanning because the rich texture features in the background needed for camera calibration often makes it hard to segment the thin structure in the foreground for reconstruction. Therefore, it would be ideal not to separate the step of camera calibration from object reconstruction, and also be able to accommodate a texture-less background for the ease of segmenting the object in the foreground. 


Another class of 3D scanning methods uses RGB-D depth sensors~\cite{Dai:tog:2017,Newcombe:ISMAR:2011}.
Most of them align and integrate depth scans using a truncated signed distance field (TSDF) representation, from which a final surface can be extracted. These methods successfully scan relatively large structures and environments. However, because of high noise and low resolution of most depth cameras, and because of the limited discretization resolution of TSDFs (e.g., on voxel grids), they fail to capture thin structures. 
Liu~\etal~\shortcite{Liu:sigga:2018} presented a new fusion primitive, curve skeletons, tailored to reliable thin structure scanning with depth cameras. However, infrared-based active RGB-D cameras are not applicable under all scene conditions, such as in strong light outdoors or when the object has black surface color. 


In this work, we present the first approach that simultaneously estimates camera poses, as well as geometry and structure of complex thin 3D structures from handheld RGB video. See Figure~\ref{fig:teaser} for some thin structures reconstructed by our method. 
%
%
%
%
%
Our method uses a new iterative optimization scheme to compute camera poses and 3D geometry and structure in a progressive manner. It first extracts 2D curves in each input video frame. After proper initialization, we solve for the camera poses and the 3D curve network by minimizing the total distance between the projections of the curve networks in each input view and their corresponding curves extracted from the same input images. Note that we add the input view one by one progressively.
As more frames are processed, the camera poses are updated, and the 3D curve network is refined in an alternating manner. In this process, we retain the junctions and the connection of points to recover the structure of the reconstruction.  Note that our method estimates camera poses solely based on thin structure objects, i.e., without requiring textured backgrounds. 



%

To summarize, we propose a novel method for reconstructing thin structures with an RGB video as input. Our approach has two main technical contributions.
\begin{itemize}
    \item We develop a new  curve-based method for efficiently and accurately computing camera parameters. This method automatically establishes the correspondence between curves in different image frames, without requiring any initial camera poses or assuming the presence of point features or texture features. This solution to the problem of automatic camera pose estimation from curves is significant in its own right. 
    \item Equipped with this camera pose estimation technique, we design an effective iterative optimization method for reconstructing 3D wire models based on a point cloud representation.  This method achieves high-fidelity reconstruction of very complex 3D wire models that have previously not been possible using a single commodity RGB camera.
\end{itemize}

\section{Related Work}
\label{sec:related}

\subsection{Thin Structure Reconstruction}
%
%
The rapid development of scanner technology 
(\eg structured light, LiDAR, and commodity RGBD camera) has motivated a large body of work on surface scanning and reconstruction (see~\cite{Berger:report:2014} for a comprehensive survey).
While most methods work on {\em extended surfaces}, some methods relax this assumption to enable the reconstruction of {\em thin surfaces}~\cite{Aroudj:sigga:2017,Savinov:cvpr:2016,Ummenhofer:iccv:2013}.
%
%
Although previous reconstruction methods have shown impressive results on smooth and textured extended surfaces, they are fragile when dealing with the thin structures that lack sufficient surface details.
Moreover, scanning thin structures proves a challenging task even with advanced acquisition technologies, particularly due to limited sensor resolution~\cite{Yan:sigga:2014,Wu:sigga:2014,Fan:sigga:2016}.

%
%
\paragraph{\bf{Reconstructing delicate structures}}
For reconstructing objects made of delicate thin structures, Li~\etal~\shortcite{li:sigga:2010} propose to reconstruct objects from high-quality 3D scans using a deformable model named {\em arterial snakes}.
An alternative approach reconstructs the surface by fitting generalized cylinders to input image~\cite{Chen:sigga:2013} or point cloud data~\cite{Yin:sigga:2014} where the fitting process is either defined manually on the 2D image plane when the skeletal curve has a simple shape or based on a 3D point cloud~\cite{Huang:sigg:2013}.
However, for the thin structures with small radius, it is extremely difficult to perform extrusion to reconstruct the curve surface as in the 3-Sweep~\cite{Chen:sigga:2013}.
\cite{Yin:sigga:2014} requires heavy user interactions and often fails to reconstruct complex thin structures when the initial 3D curve skeleton extracted by L1-axis \cite{Huang:sigg:2013} contains topological and geometric errors.




%
%
\paragraph{\bf{Image-based reconstruction}}
Another line of works reconstructs thin structures from multiple color images.
%
%
Tabb~\shortcite{Tabb:cvpr:2013} reconstructs thin structures from multiple images using silhouette probability maps by solving a pseudo-Boolean optimization problem in a volumetric representation. The capturing system they used consists of 30 cameras, including 20 industrial-grade cameras, that are fixed on walls and ceilings surrounding the object to be scanned. 
%
\cite{Tabb:IROS:2017} proposes a method that uses a robotic vision system for image acquisition to reconstruct tree branches in a real field outdoor. Both of these two methods require pre-computed camera poses as input.
Martin~\etal~\shortcite{Martin:sigga:tb:2014} propose to reconstruct thin tubular structures (\eg cables) using physics-based simulation of rods. A special voxel grid, called occupancy grid, is used to resolve crossing ambiguities observed in the 2D image plane.
Hsiao~\etal~\shortcite{Hsiao:sigga:2018} reconstruct a 3D wire art from several views of 2D line-drawings. This method uses constrained 3D path finding in a volumetric grid to resolve spatial ambiguities due to inconsistency between input line drawings. They demonstrate impressive results for a wide set of wire art models, but the voxel representation precludes reconstruction at very high accuracy.
%
Li~\etal~\shortcite{li:cvpr:2018} use spatial curves generated from image edges to reconstruct thin structures.  
Liu~\etal~\shortcite{Liu:sigg:2017} reconstruct 3D wire models from a small number of images using a candidate selection strategy for image correspondence. 
While these image-based methods produce impressive reconstructions of wire models of moderate complexity, they suffer from the error-prone camera pose prediction and their inability to handle self-occlusion of 3D wires. They thus perform poorly when reconstructing complex 3D thin structures.
Y\"{u}cer~\etal~\shortcite{Yucer:3dv:2016} exploit the local gradient information in captured dense light fields to segment out thin structures~\cite{Yucer:tog:2016} and estimate a per-view depth map for voxel-based carving and Poisson surface reconstruction~\cite{Kazhdan:sgp:2006}.
This method assumes that input objects have sufficient texture details for a valid depth estimation.
In contrast, our method employs a curve-based method for accurate camera pose estimation and effectively handles textureless objects and self-occlusion to achieve high-fidelity reconstruction of complex 3D thin structures from a handheld color video, which has not been possible before.

\begin{figure*}[!t]
    \centering
    \includegraphics[width=\linewidth]{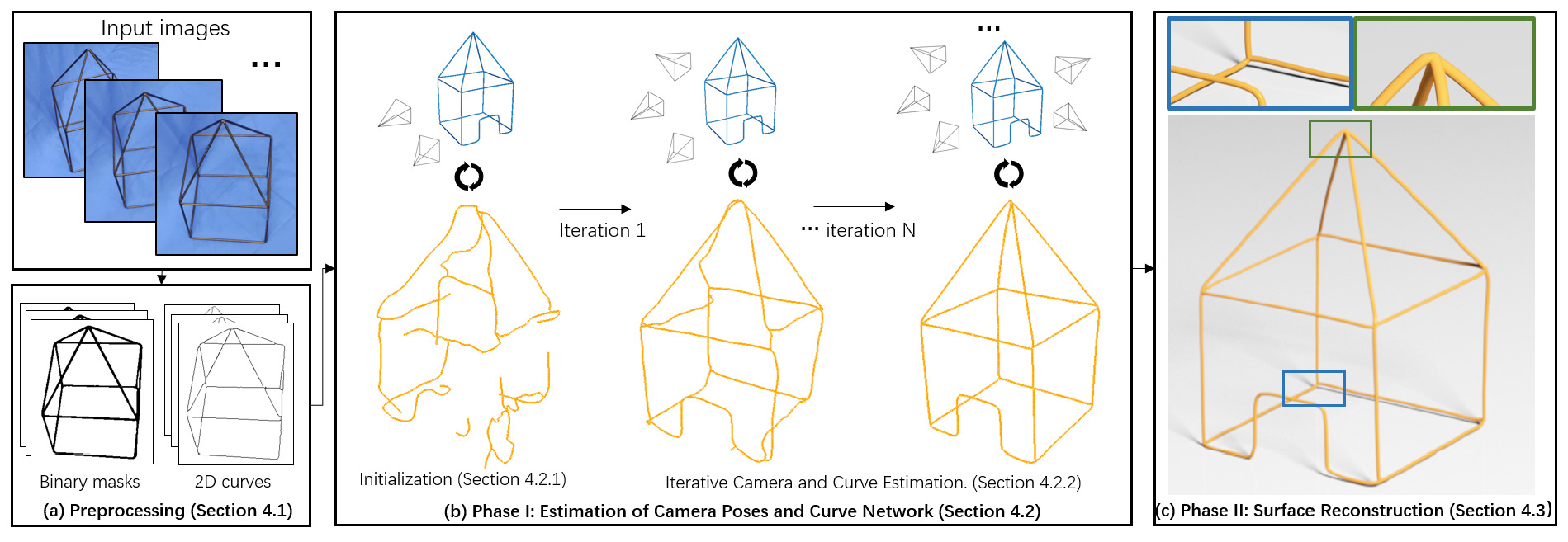}
    \caption{
    \tb{Method overview.}
    Given a sequence of RGB images of a 3D thin structure, we first segment out the structure in foreground to obtain a sequence of binary masks and corresponding one-pixel wide 2D curves in the preprocessing step (a). To solve the optimization problem formulated in~\secref{phaseI}, we first initialize the camera poses and a 3D curve network using two properly selected image frames from the input video (\secref{initialize}). (b) Then we adopt an iterative structure optimization strategy that adds and processes the remaining image frames progressively to update the camera poses of observed views so far and refine the estimated curve network in an alternating manner (\secref{phaseI}). (c) The surface of the thin structure is modeled as a sweep surface along the constructed 3D curve with a circular section whose radius estimated from image observations (\secref{surf_reconstruct}). The final reconstructed output is a clean, smooth thin structure with high geometric and topological fidelity to the original wire model.
    }
    \label{fig:overview}
\end{figure*}

%
%
\paragraph{\bf{Depth-based reconstruction}}
%
Several approaches use 
depth sensors for thin structure reconstruction. 
They can, therefore, resort to actively measured depth rather than difficultly matched image correspondence to infer 3D structure.
However, depth camera data come with their own specific challenges, as the measured depth is often of low resolution, very noisy, and often exhibits systematic measurement biases. 
Depth-based reconstruction algorithms need to address these challenges specifically.

%
%
%
Liu~\etal~\shortcite{Liu:sigga:2018} present {\em CurveFusion} that directly explores a dense 3D point cloud obtained from commodity depth sensors and utilizes curve skeleton as a fusion primitive for thin structure reconstruction given the aforementioned depth data challenges.
The infrared-based depth sensors used in this method make it applicable only in some scene conditions, such as indoor environment without strong sunlight and objects with non-black surface color, which significantly limits its application.
Moreover, the maximum reconstruction accuracy is bounded by the limited depth resolution of depth sensors, which often leads to missing curves or incorrect topology on the complex junctions. 
In contrast, our method works directly on a higher resolution RGB video to establish the correspondence using a curve-based matching approach. Therefore it can reconstruct thin structures with higher spatial resolution.

\subsection{Curve-based Structure-from-Motion}
\label{sec:curveSFM}
Our core algorithm builds upon the concepts of 3D reconstruction using structure-from-motion (SfM)~\cite{Snavely:sigg:2006,schonberger:cvpr:2016} and multi-view stereo (MVS)~\cite{Goesele:iccv:2007,Furukawa:pami:2010,schonberger:eccv:2016,kuhn:ijcv:2017,huang:cvpr:2018}.
These methods follow a general principle to establish point correspondences across images and then reconstruct a sparse set of 3D points alongside camera poses.
While impressive results were achieved in reconstructing objects with reliable textured surfaces, these methods perform poorly when there are insufficient easy-to-discriminate point features in the scene.
Therefore, several approaches exploit higher order features such as lines~\cite{Jain:cvpr:2010,hofer:3dv:2014} or curves~\cite{Xiao:josaa:05,Fabbri:cvpr:2010,Nurutdinova:iccv:2015,Usumezbas:eccv:16} as reconstruction primitives.
We refer the reader to the work by Fabbri~\etal~\shortcite{Fabbri:ijcv:2016} for a fundamental theory of the multi-view geometry on curves. 

Existing curve-based reconstruction methods can be classified according to the assumptions they made.
For instance, Berthilsson~\etal~\shortcite{Berthilsson:ijcv:2001} assume that the reconstructed 3D curve should be occlusion-free in each view.
Hong~\etal~\shortcite{Hong:eccv:2004} focus on reflective and symmetric curves.
Rao~\etal~\shortcite{Rao:irs:2012} present a curve-based SLAM method that is closely related to our method in the spirit of jointly estimating camera poses and thin structures from curve primitives.
However, their method assumes the input of stereo images and the curves of interest are constrained to planar curves on the ground, while our method takes input images from a hand-held RGB camera and works for general 3D spatial curve networks in the setting of thin structure reconstruction. 
%
%
%
Similarly, Nurutdinova~\etal~\shortcite{Nurutdinova:iccv:2015} assume that the correspondences between image curves are given and use curve as primitive to improve the accuracy of SfM.
%
%
Recent works~\cite{Fabbri:cvpr:2010,Usumezbas:eccv:16} relax the assumptions and follow a typical workflow that reconstructs 3D curves by aggregating detected 2D curve segments across images using epipolar constraints.
All the above methods share the limitation that they consider each line or curve segment individually, which possibly suffer from noise and reconstruction gaps, and do not reason about the global curve structure and connectivity of objects.
In contrast, our method estimates camera poses and reconstructs a continuous 3D curve with an effective measure to handle self-occlusion. It produces faithful reconstruction of the complex thin structures and their connectivity way beyond the capability of existing methods.

%
%
%

\section{Overview}
\label{sec:overview}
Our goal is to extract the 3D geometry as well as structure and topology of objects composed of complex networks of thin tubular-shaped or wire-like structures of varying radius from a handheld color video.  
%
While we assume the intrinsic parameters of the camera to be known and fixed, the motion (i.e., the sequence of poses) of the camera is unknown and needs to be computed. 
We follow the definition used in previous approaches~\cite{Liu:sigg:2017,Liu:sigga:2018} to define a thin structure as a collection of connected generalized cylinders, where each generalized cylinder is a sweeping surface of varying radius along its {\em skeletal curve}. 
Therefore, we represent a thin structure as a network of connected skeletal curves, to be called the {\em curve network} for short, with a radius function defined on the skeletal curves.
\figref{overview} illustrates the pipeline of our method.
Given a sequence of RGB images, we first segment out the thin structure in the foreground to obtain a sequence of binary masks and corresponding one-pixel-width 2D curves in a preprocessing step (see~\figref{overview}(a), \secref{preprocess}).
Our method then runs in two phases: {\em Phase I - camera pose estimation and curve network reconstruction}; and {\em Phase II - surface geometry reconstruction}.
%
Phase I computes two sets of variables in an iterative optimization approach (\secref{phaseI}): 
the set of extrinsic parameters defining the camera poses of all input frames, and the set of 3D points defining the curve network in 3D, along with accurate connectivity classification at junctions of the curve network. 
After an initialization step (\secref{initialize}), we adopt an iterative structure optimization to compute the values of the aforementioned variables by minimizing the difference between the projection of the curve network and the corresponding 2D curves observed in all input image frames (\secref{iteration}).
Frames of the input video are progressively added and processed in their temporal order.
When a new frame is added, a new set of 3D points and refined camera poses based on all images seen so far is estimated in an iterative optimization process that alternates between camera pose and 3D geometry computation (see~\figref{overview}(b)).
%

To enable this, we developed two techniques that are critical to the success of our method. 
The first technique is a new correspondence finding algorithm that succeeds on textureless thin objects by operating on curve features instead of salient keypoints. 
It is designed to reliably establish correspondences between the estimated 3D curve network and observed 2D curves.
The second is an efficient and effective method to detect self-occlusion of thin structures in input views. It enables us to prevent unreliable 2D curve segments observed from a self-occluding perspective from being considered in the update of the corresponding 3D curve points.
This significantly improves the quality of the final reconstructed curve network (\secref{occlusion}).

Once all frames are processed, Phase II reconstructs the final surface of the thin structure by sweeping along skeletal curves a disc whose varying radius (or thickness) is estimated by fusing the 2D curve radius estimates across all the input image observations (see~\figref{overview}(c), \secref{surf_reconstruct}).

\section{Algorithm}
\label{sec:algo}

\subsection{Preprocessing}
\label{sec:preprocess}
During preprocessing, we first segment each input image to obtain a binary mask of the pixels showing the thin structure in the foreground. Reliable segmentation of thin structures from a general background is extremely challenging. 
We assume that the 3D thin structure is filmed in front of a simple uniform background so it can be segmented using color keying method or advanced video object segmentation tool such as Rotobrush in Adobe After Effects~\cite{Bai:sigg:2009}. For the shapes of small to medium size, this can be easily achieved by filming in front of a monochromatic wall or cloth. 
Let $\inputI=\{\inputIk{k}, k=1,2,...,\frameNum\}$ denote the input video as a sequence of RGB frames, and let $\segI=\{\segIk{k}, k=1,2,...,\frameNum\}$ denote the corresponding foreground segmentation masks.
We use the image thinning method to extract one pixel wide medial axis curves from $\segI$, which are called {\em skeletal curves} and henceforth denoted by $\skelC=\{\skelCk{k} \subset  \mathbb{R}^2, k=1,2,...,\frameNum\}$.
%
%
%

\subsection{Phase I: Estimation of Camera Poses and Curve Network}
\label{sec:phaseI}
%
We develop an optimization framework to simultaneously compute all camera poses and a 3D skeletal curve network.
The camera pose of input image $\inputIk{k}$ is parameterized by the respective camera rotation $\camRk{k}$ and translation $\camTk{k}$.
%
The 3D curve network $\curveStruct$ of the entire object to be reconstructed is represented as a graph $G=(\mathbf{P},\mathbf{E})$, where $\mathbf{P}=\{P_i \in \mathbb{R}^3, i=1,2,...,\curvePntNum\}$ represents a sequence of 3D points appropriately sampled on the 3D curve and the edge set $\mathbf{E}$ encodes the connectivity of the 3D points.
%



%
We compute the camera poses, $(\camRk{k}, \camTk{k})$, and the curve network, $\curveStruct$, by minimizing an objective function that measures the sum of the squared 2D distances between the projection of the curve network, $\curveStruct$, in an input view $\inputIk{k}$ and the corresponding 2D skeletal curve $\skelCk{k}$, across all input views: 
%
\begin{equation}
    F(\{\camRk{k}, \camTk{k}\}; \curveStruct) = \sum_{k}{\rm dist}^2(\curveProjk{k}, \skelCk{k}),
    \label{eq:obj_function}
\end{equation}
where $\curveProjk{k}=\pi (\camRk{k}, \camTk{k}; \curveStruct)$ is the projection of $\curveStruct$ onto view $\inputIk{k}$ using camera extrinsic parameters $(\camRk{k}, \camTk{k})$. 
%
Further, ${\rm dist}^2(\curveProjk{k}, \skelCk{k})$ indicates the one-sided integrated squared distance between the curve $\curveProjk{k}$ and the curve $\skelCk{k}$ in the 2D image plane, which is defined as: 
\[
{\rm dist}^2(\curveProjk{k}, \skelCk{k})= \int_{p \in \curveProjk{k}} \min_{q \in \skelCk{k}} \|p-q\|_2^2,
\]
where $\|\cdot \|_2$ is the Euclidean norm.
In the following, we explain an iterative scheme to minimize this objective function.
Starting from two input images, our method progressively adds input frames one-by-one to a set of active images to estimate $(\camRk{k}, \camTk{k})$, and refines $\curveStruct$ on the currently active image set in an alternating manner.


\subsubsection{Initialization}
\label{sec:initialize}
%

Now we discuss how to initialize the camera poses and the 3D curve network in order to solve the optimization problem formulated in~\eqnref{obj_function}.
Since a proper initialization will largely facilitate the later incremental reconstruction of thin structure, we propose to select two image frames that meet the criterion of presenting sufficient camera movement such that the resulting parallax could provide sufficient depth information.
Specifically, we prepare a sequence of image pairs, denoted as $\{(\inputIk{1}, \inputIk{i}), i=2, 3, \dots \}$ and evaluate the image pairs successively in the following way. 
Given an image pair, we compute a pixel-wise correspondence between the curve segments using a 2D curve matching method~\cite{kraevoy:eg:2009} that exploits optical flow~\cite{eccv:kroeger:2016} to estimate an initial coarse alignment between two images.
Based on the correspondence induced by the matching curves, we jointly estimate the camera poses of these two frames and a 3D point cloud using the bundle adjustment~\cite{triggs:bundle:1999}.
We then choose the first pair where the distance between the two initially estimated camera locations is larger than a threshold $\beta$ ($\beta = 0.03$ in our experiments). Note that the distance is normalized by the average depth of the points relative to the first camera.

\paragraph{Curve network generation}
After the initial 3D point set is generated, we connect the points into a 3D curve network using a variant of Kruskal's algorithm to compute the minimum spanning tree~\cite{kruskal:shortest:1956}.
The goal is to build a general curve structure rather than a tree-like structure for representing a wide range of object.  
%
%
The network formation algorithm has the following steps:
\begin{enumerate}
\item  Given a set of 3D points, $\mathbf{P}$, we first form a set of candidate edges $\mathbf{E'}$ by pairing points with a mutual distance below 
a preset threshold $\theta$, that is, 
$\mathbf{E'} =\{(P_i, P_j) | P_i, P_j \in \mathbf{P},  i\neq j,  \|P_i - P_j \|< \theta )\}$. 
We empirically set $\theta = 5 \delta_0$, where $\delta_0$ is the sampling distance defined below.
\item  A graph $G=(\mathbf{V},\mathbf{E})$ is maintained, with $\mathbf{V}=\mathbf{P}$ and $\mathbf{E}=\emptyset$ (empty set). 
\item Check, in ascending order of the edge length, if the edges of $\mathbf{E'}$ shall be added to $\mathbf{E}$.
%
An edge $(P_i, P_j) \in \mathbf{E'}$ is added to $\mathbf{E}$ if it does not form a loop in the graph $G$ or the length of the loop formed by the edge is greater than a threshold $L$. We set $L = 20 \delta_0$ in our experiments, where $\delta_0$ is the sampling distance defined below.
\item Repeat step (3) until all edges in $\mathbf{E'}$ have been processed.
\end{enumerate}

However, only loops larger than a minimal size threshold shall be allowed in the curve network in order to avoid the formation of erroneous small loops due to noise in the point cloud.

\paragraph{Curve re-sampling}
Since the 3D points in $\mathbf{P}$ that define the geometry of curve network $G$ are directly "lifted" from pixels in an input view, the distribution of the points is in general not uniform.
We improve this by re-sampling points on the curve network with an even distribution. 
%
%
We suppose that the center of the object being scanned has unit distance to the camera center. Then the inter-point sampling distance is set to be $\delta_0 = 1/f_0$, where $f_0$ is the camera focus length, which is identical across all the input frames. The rationale for setting the sampling distance in this way is to ensure that, when projected onto the input view, the sample points on the 3D curve have nearly the same density as the image pixels.

Note that both aforementioned steps, curve network generation and curve re-sampling, are performed not only during initialization but also when the points are updated in every iteration of minimizing the objective function defined in~\eqnref{obj_function}.



\subsubsection{Iterative Camera and Curve Estimation}
\label{sec:iteration}
\begin{figure}[!t]
    \mpage{0.33}{\includegraphics[width=\linewidth]{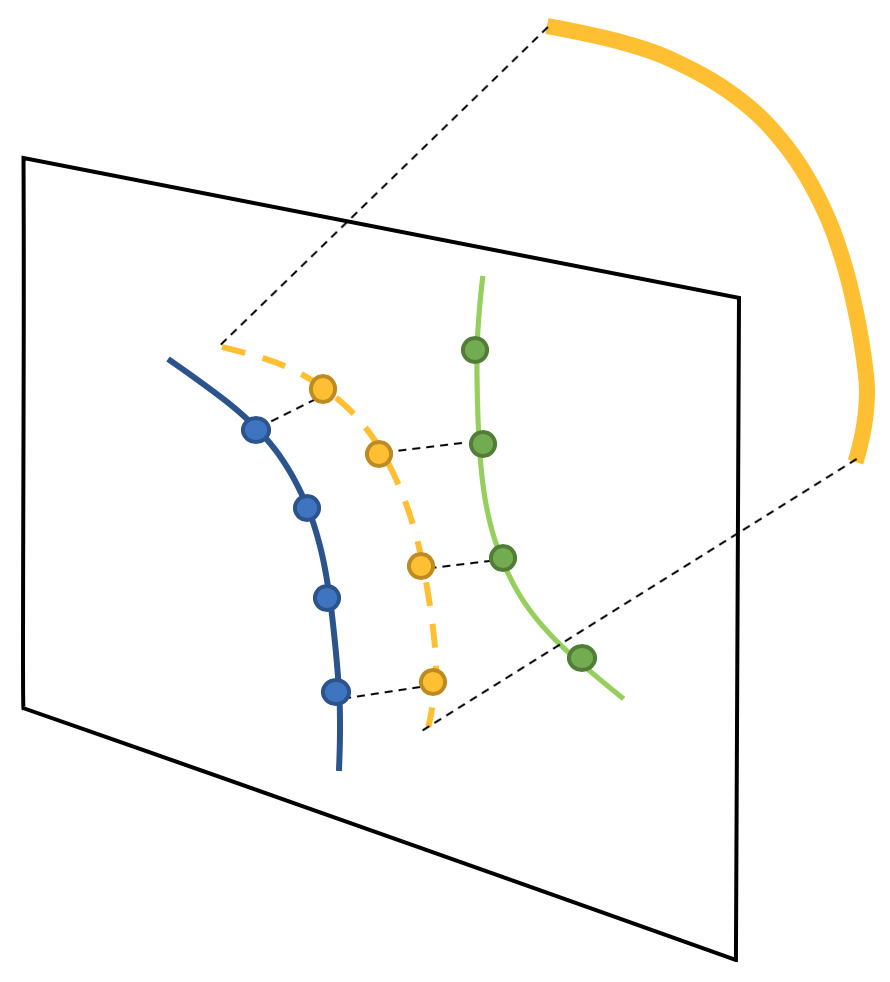}}
    \mpage{0.3}{\includegraphics[width=\linewidth]{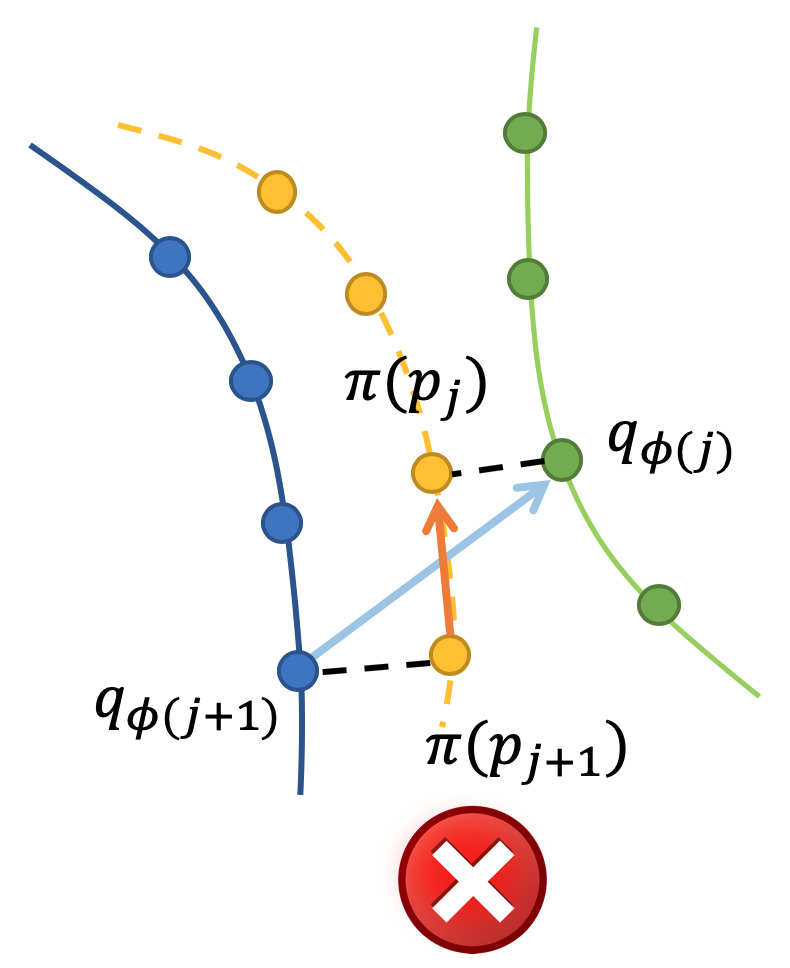}}
    \mpage{0.3}{\includegraphics[width=\linewidth]{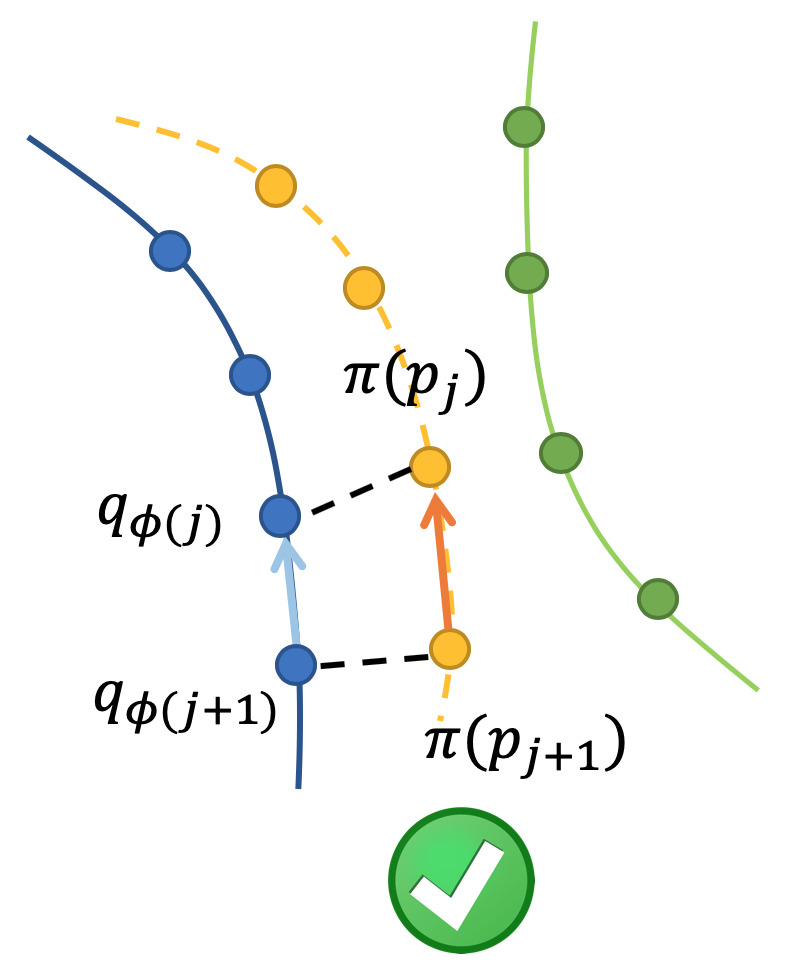}}
    \\
    \mpage{0.33}{(a)}
    \mpage{0.3}{(b)}
    \mpage{0.3}{(c)}
    
    \caption{\tb{Illustration of curve matching.} (a) A naive closest point method may lead to (b) inconsistency matching between the projection of 3D curve and the 2D curves; (c) Correct matching by considering both distance and curve consistency.
    }
    \label{fig:matching}
\end{figure}

\paragraph{Curve matching.}
%
A key step in solving the minimization problem in~\eqnref{obj_function} is our {\em curve matching} step that establishes the correspondence between the points of 3D curve $\curveStruct$ and those on the observed image curves $\skelCk{k}$ in each input image.
%
A naive distance-based criterion would be finding the closest point $q_j \in \skelCk{k}$ to the projection of any given point $P_j \in \curveStruct$, and setting $q_j$ as the matching point of $P_j$.
Clearly, this greedy approach may often yield wrong matching points when the observed image curve $\skelCk{k}$ is dense and cluttered (see~\figref{matching}(a)).

We solve this problem by combining a distance-based criterion with a constraint on curve consistency, which indicates that two consecutive sample points $\{P_i,P_{i+1} \in \curveStruct\}$ should have their corresponding points $\{q_j,q_{j+1}\}$ lie on the same local branch of the image curve $\skelCk{k}$ and be close to each other (see~\figref{matching}(b,c)).
%

%
Specifically, suppose that the 3D curve $\curveStruct$ is represented by the point sequence $\mathbf{P}=\{P_i \in \mathbb{R}^3, i=1,2,..., \curvePntNum\}$ and the observed image curve $\skelCk{k}$ in the image $\inputIk{k}$ is represented by the pixel sequence $\mathbf{Q}=\{q_i \in \mathbb{R}^2, i=1,2,..., \skelPntNum\}$. A matching between $\mathbf{P}$ and $\mathbf{Q}$ is then defined by an injection $\phi$ from $[1, \curvePntNum]$ into $[1, \skelPntNum]$. Let $\Phi$ denote the set of all such injections. Then we seek the optimal matching $\phi_0$ that minimizes an objective function defined as follows:
\begin{equation}
\begin{aligned}
E_{match} (\curveStruct, \skelCk{k}) = \alpha \sum_{j} \|\pi (P_j) - q_{\phi(j)}\| + \\ \sum_{j} \|(\pi(P_j) - \pi(P_{j+1})) - (q_{\phi(j)} - q_{\phi(j+1)})\|.
\end{aligned}
\end{equation}
We empirically set $\alpha=0.1$ in our experiments.
This minimization problem is solved efficiently with dynamic programming. 
Given a point $P_i \in \curveStruct$, we select the candidates for $P_j$'s matching points only among those pixels $q_j \in \skelCk{k}$ lying within a circle of radius of $10$ pixels centered at the projected point $\pi (P_i)$.


\paragraph{Optimization strategy}
There are two sets of variables in the objective function (see~\eqnref{obj_function}): (1) the camera pose parameters $(\camRk{k}, \camTk{k})$ and (2) the sample points of the curve network $\curveStruct$.
To simplify the optimization task, we process the camera poses in a progressive manner so that a new camera view is added after the camera poses of all the preceding views have been estimated. 
Similarly, the curve network is also refined in a progressive manner.

Based on the discrete representation of the curve network, we can rewrite the~\eqnref{obj_function} as:
\begin{equation}
    \tilde{F}(\{\camRk{k}, \camTk{k}\}; \curveStruct) = \sum_{k}  \sum_{P_j \in \curveStruct} \occluFunc(P_j, \inputIk{k}) \cdot 
    e_{k,j}
    \label{eq:obj_function_discrete}
\end{equation}
where the distance error term
\begin{equation}
    e_{k,j} = \min_{q \in \skelCk{k}} \|\pi (\camRk{k}P_j + \camTk{k}) - q\|_2^2
    \label{eq:dist_error}
\end{equation}
is the squared distance from the projection of a variable point, denoted as $p'_j = \pi (\camRk{k}P_j + \camTk{k})$, to the observed image curve $\skelCk{k}$.
And $\occluFunc(P_j, \inputIk{k})$ is an indicator function that returns value 0 if the point $P_j$ fails to pass the self-occlusion test with respect to the input view $\inputIk{k}$, and otherwise returns value 1.
The details of self-occlusion handling process are elaborated in~\secref{occlusion}.

\begin{figure}[!t]
    \mpage{0.45}{\includegraphics[width=\linewidth]{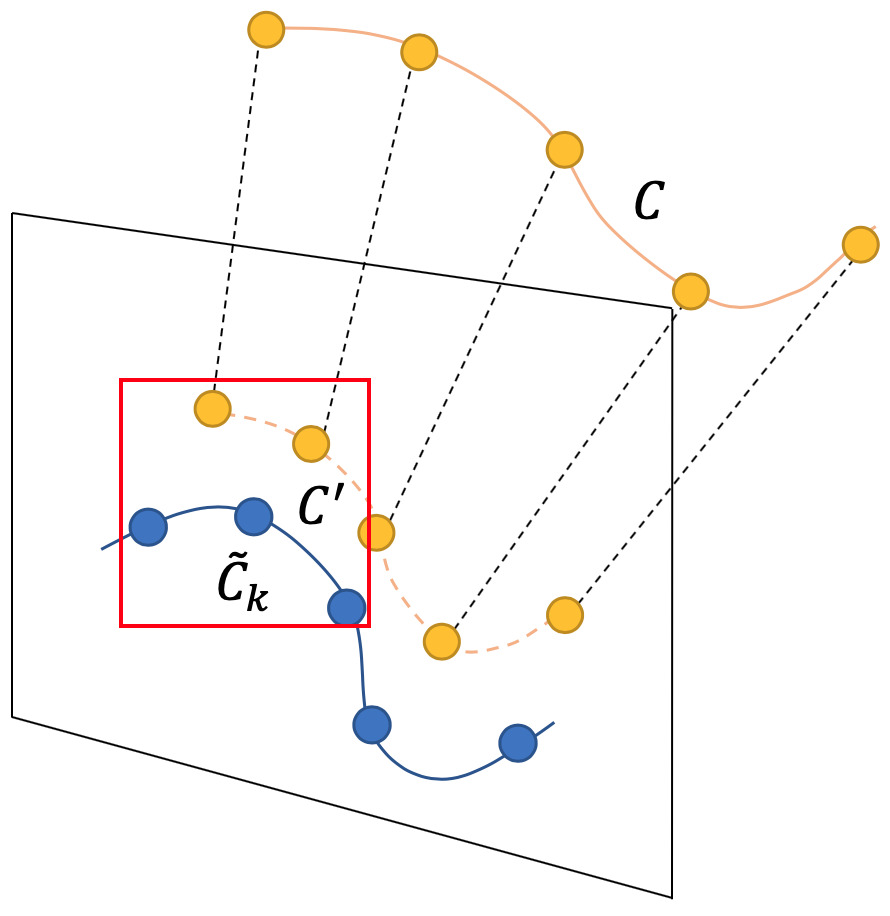}}
    \mpage{0.45}{\includegraphics[width=\linewidth]{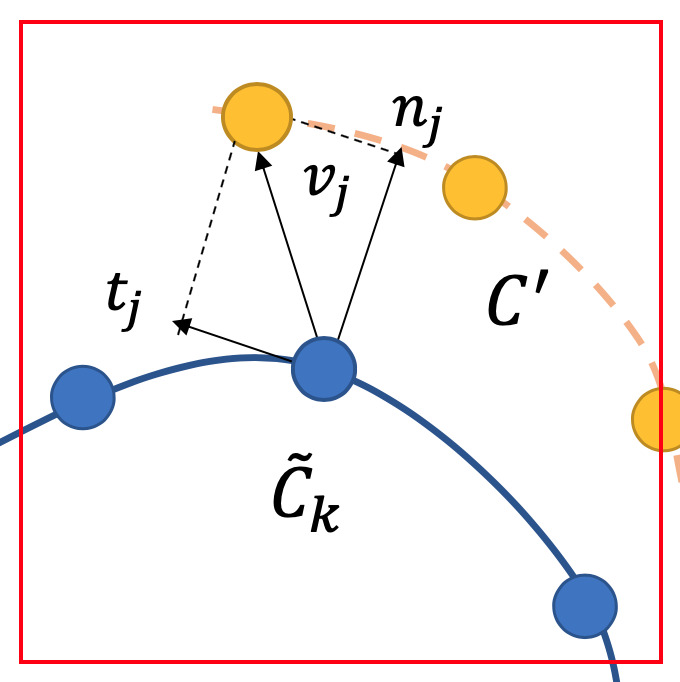}}
    \\
    \mpage{0.45}{(a)}
    \mpage{0.45}{(b)}
    
    \caption{(a) The projection $\curveProjk{}$ of 3D curve $\curveStruct$ and its corresponding image curve $\skelCk{k}$; (b) The distance error term $ e_{k,j}$ is expressed as a linear combination of squared distances along the tangent direction $t_j$ and the normal direction $n_j$, respectively.
    }
    \label{fig:error_term}
\end{figure}

\begin{figure}[!b]
    \mpage{0.55}{\includegraphics[width=\linewidth]{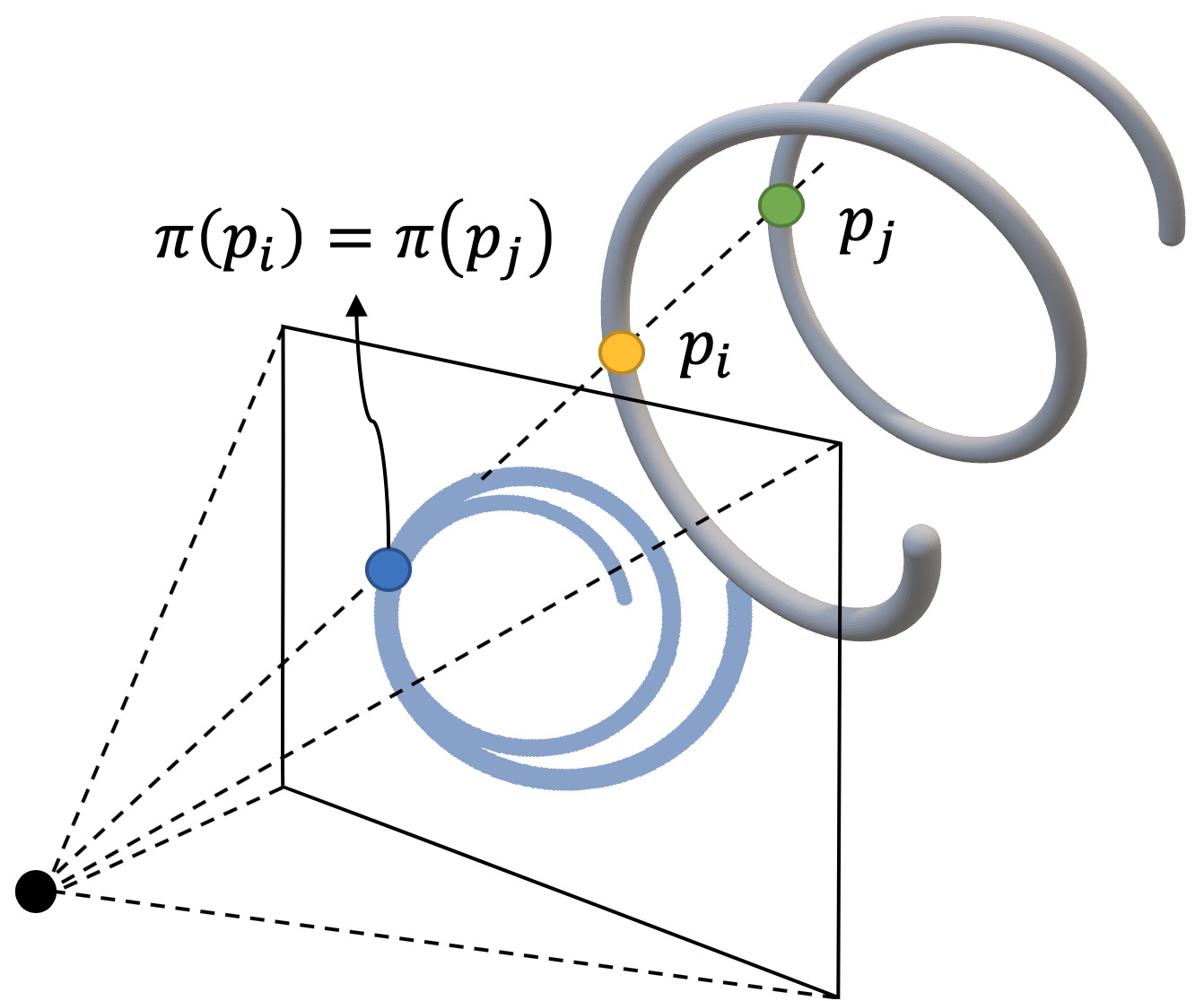}}
    \mpage{0.35}{\includegraphics[width=\linewidth]{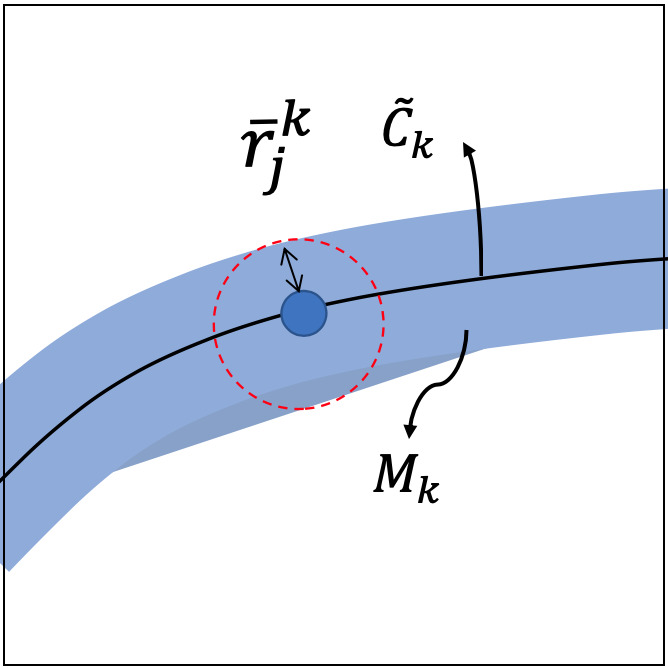}}
    \\
    \mpage{0.55}{(a)}
    \mpage{0.35}{(b)}
    
    \caption{ (a) Illustration of self-occlusion: Two distinct points, $p_i$ and $p_j$, on an non-self-intersecting 3D curve $\curveStruct$ are projected to the same intersecting point $\pi (p_i)=\pi (p_j)$ in the view plane; (b) The width $2\bar{r}_j^k$ of the projected point of 3D thin structure is estimated by the corresponding 2D curve $\skelCk{k}$ and segmentation mask $\segIk{k}$ in the view of $\inputIk{k}$.
    }
    \label{fig:self_occlusion}
\end{figure}

\begin{figure*}[!t]
    \includegraphics[width=\linewidth]{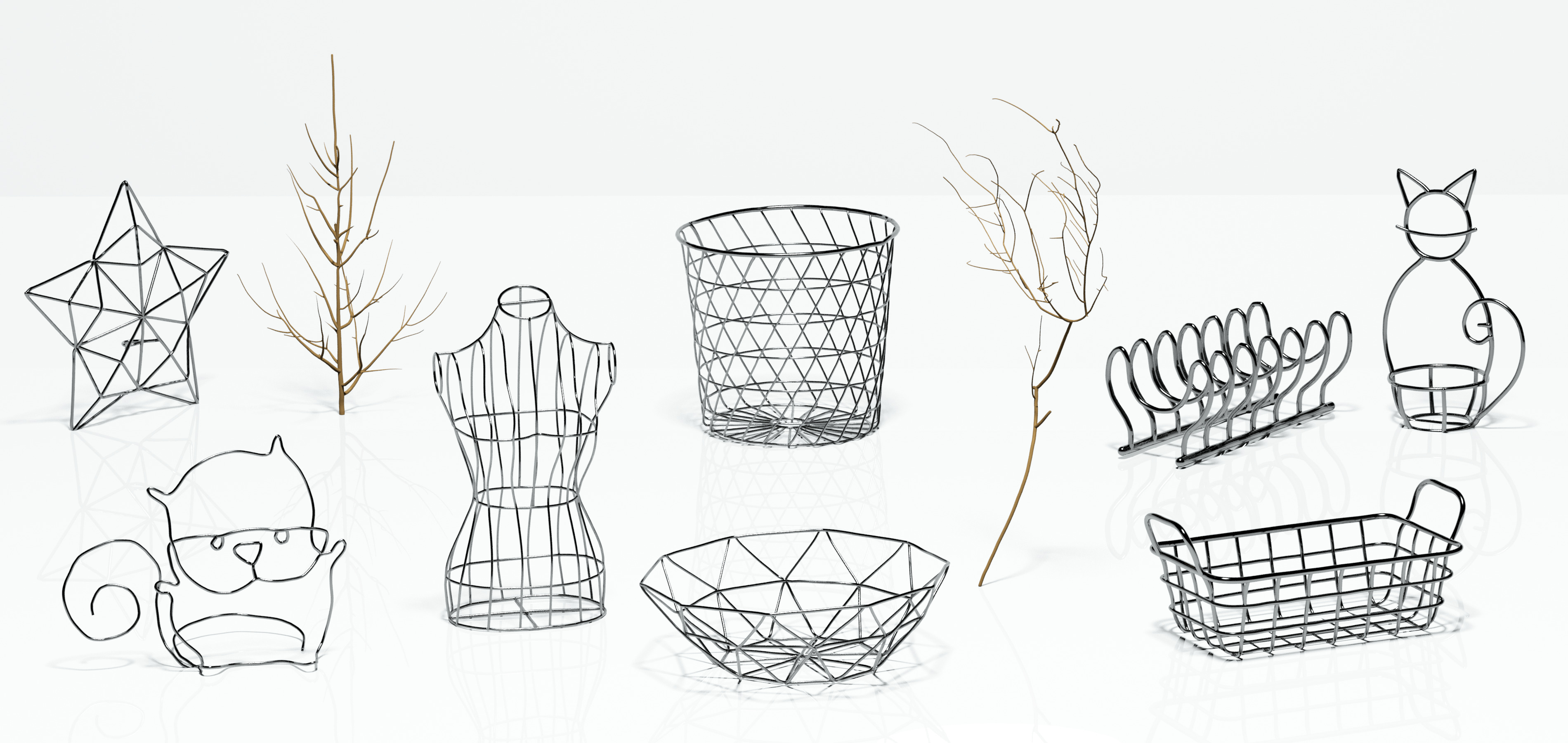}
    \caption{\tb{Reconstruction results.} 
    The gallery of real world 3D thin structures reconstructed using our method. Our method reconstructs a wide variety of wire objects in high quality.
    }
    \label{fig:visual_results}
\end{figure*}

In practice, we adopt a commonly used formulation for curve fitting~\cite{pottmann:CGA:2002} to efficiently minimize the distance error term $e_{k,j}$ in~\eqnref{dist_error} by expressing it as a special linear combination of the squared distances of $p'_j$ to the image curve $\skelCk{k}$ along the normal direction, $n_j$, and tangent direction, $t_j$, at the corresponding point $q_j \in \skelCk{k}$ (see~\figref{error_term}).
Thus, by denoting $v_j = \pi (\camRk{k}P_j + \camTk{k}) - q_j$, \eqnref{obj_function_discrete} can be further expressed as:
%
\begin{equation}
     \tilde{F^*} (\{\camRk{k}, \camTk{k}\}; \curveStruct) = 
     \sum_{k}  \sum_{P_j \in \curveStruct} \occluFunc(P_j, \inputIk{k}) 
     \cdot e^*_{k,j}
     + F_s(C)
    \label{eq:obj_function_final}
\end{equation}
%
where the distance error term
\begin{equation}
    e^*_{k,j} = \| v_j \cdot n_j \|^2 + w \| v_j \cdot t_j\|^2
    \label{eq:dist_error_slide}
\end{equation}
and the regularization term
\begin{equation}
F_s(C) = \lambda \sum_{k}  \sum_{P_j \in \curveStruct} \| P_{j+1} - 2 P_j + P_{j-1}\|^2
\end{equation}
We set the weight parameter $\lambda = (\frac{2.5}{\delta_0})^2$ and $w$ with value $0.5$ in our experiments for a trade-off between efficiency and stability of convergence.

We minimize the objective function $\tilde F^*$ iteratively. 
In each iteration, $\tilde F^*$ is minimized in an alternating fashion that first optimizes the camera poses $\{(\camRk{k}, \camTk{k})\}$ while fixing the curve points $\{p_j\}$, and then optimizes the curve points while fixing the camera poses. The process iterates until the curve points stop moving.
Note that, once the camera poses $\{(\camRk{k}, \camTk{k})\}$ and the points $\{P_j\}$ have been updated, we re-compute the corresponding point $q_j$ on the image curve $\skelCk{k}$ to the updated projected point $p'_j$ by our matching algorithm before entering the next iteration.

\subsubsection{Handling Self-occlusion} 
\label{sec:occlusion}

A {\em self-occlusion} occurs when distinct parts of a 3D curve network $\curveStruct$ are projected onto the same location in a 2D view $\inputIk{k}$. 
When this happens, the 3D curve $\curveStruct$ intersects itself in the projected view where the intersecting pixels correspond to multiple points of $\curveStruct$, causing ambiguity during the reconstruction (see~\figref{self_occlusion}(a)).
Hence, it is critical to detect self-occlusion to ensure that image observations involving self-occlusion are not used for 3D curve reconstruction. 

To determine whether the points of 3D curve $\curveStruct$ are subject to self-occlusion in a certain image $\inputIk{k}$, we perform the following steps:
\begin{enumerate}
\item For each point $P_j \in \curveStruct$ and its matching point $q_{\phi(j)} \in \skelCk{k}$ paired in the curve matching step, we examine the neighboring pixels of $q_{\phi(j)}$ using a $3 \times 3$ local window. We then generate a 3D point set $\hat{\mathbf{P}}$ that contains whatever points of $\curveStruct$ that match the pixels within the local window.
\item We then compute a spatial compactness factor as the standard deviation $\sigma_j$ derived from the average L2 distance between the points in $\hat{\mathbf{P}}$ and their centroid.
\item Point $P_j$ is labeled as self-occluded (\ie $\occluFunc(P_j, \inputIk{k})=0$) if $\sigma_j \geq \sigma_0$. Otherwise if $\sigma < \sigma_0$ and all the points in $\hat{\mathbf{P}}$ lie on the same local branch of the 3D curve $\curveStruct$, the point is not self-occluded (\ie $\occluFunc(P_j, \inputIk{k})=1$).
\end{enumerate}
We use a preset threshold $\sigma_0=10 \delta_0$ in our experiments, where $\delta_0$ is the sampling distance defined in Sec.~\ref{sec:initialize}. 
We employ this self-occlusion handling not only during the 3D curve reconstruction but also during the surface reconstruction in Phase II in order to obtain reliable radius estimates as discussed below.

\subsection{Phase II: Surface Geometry Reconstruction}
\label{sec:surf_reconstruct}
To generate the generalized cylinder (or sweeping surface) representing the surface of the reconstructed thin structure, we compute the radius at each point of the 3D curve using the corresponding image observations from all the input views.
Specifically, for each point $P_j$ of 3D curve $\curveStruct$, we first find its matching point $q_{\phi(j)} \in \skelCk{k}$. 
Note that the projection of a generalized cylinder is a 2D strip in an input image.
Therefore we compute the width of the strip at the point $q_{\phi(j)}$ by computing the distances from $q_{\phi(j)}$ to both sides of the strip defined in the foreground segmentation mask $\segIk{k}$. We denote the estimated width by $2\bar{r}_j^k$ (see~\figref{self_occlusion}(b)).
Then the radius of the thin structure at $P_j$ in the image view $\inputIk{k}$ is defined as:
\begin{equation}
r_j^k = \frac{\bar{r}_j^k}{f_0} \cdot {\rm depth}(P_j, \inputIk{k}),
\end{equation}
where $f_0$ is the focal length of the camera and ${\rm depth}(P_j, \inputIk{k})$ is the depth of $P_j$ with respect to the image $\inputIk{k}$.
The final estimated radius at $P_j$ is thus defined as the average of the $r_j^k$ over all the input images, excluding those image observations that involve self-occlusion at $P_j$, \ie $\occluFunc(P_j, \inputIk{k})=0$.

\section{Experimental Results}
\label{sec:results}
We evaluate our method on a wide variety of real world thin structures with varying scale and complexity, \eg from man-made wire sculptures to leafless tree branches.
The input data is captured by a hand-held camera with known intrinsic parameters.
The length of each video ranges from 20 to 30 seconds, depending on the scale and complexity of the objects, and we downsample the videos with a sampling rate of 5 frames to produce $100 \sim 300$ RGB images as input frames to our method.
To facilitate the foreground segmentation, all the objects were captured against a clean background.

\figref{visual_results} shows a gallery of real world 3D thin structures reconstructed using our method.
%
We refer the reader to the supplementary video for more results, with close-ups.
These results demonstrate the effectiveness of our method in estimating accurate cameras poses as well as producing high-fidelity reconstructions of 3D thin structures with complex geometry and topology.

\begin{figure}[!t]
\includegraphics[width=\linewidth]{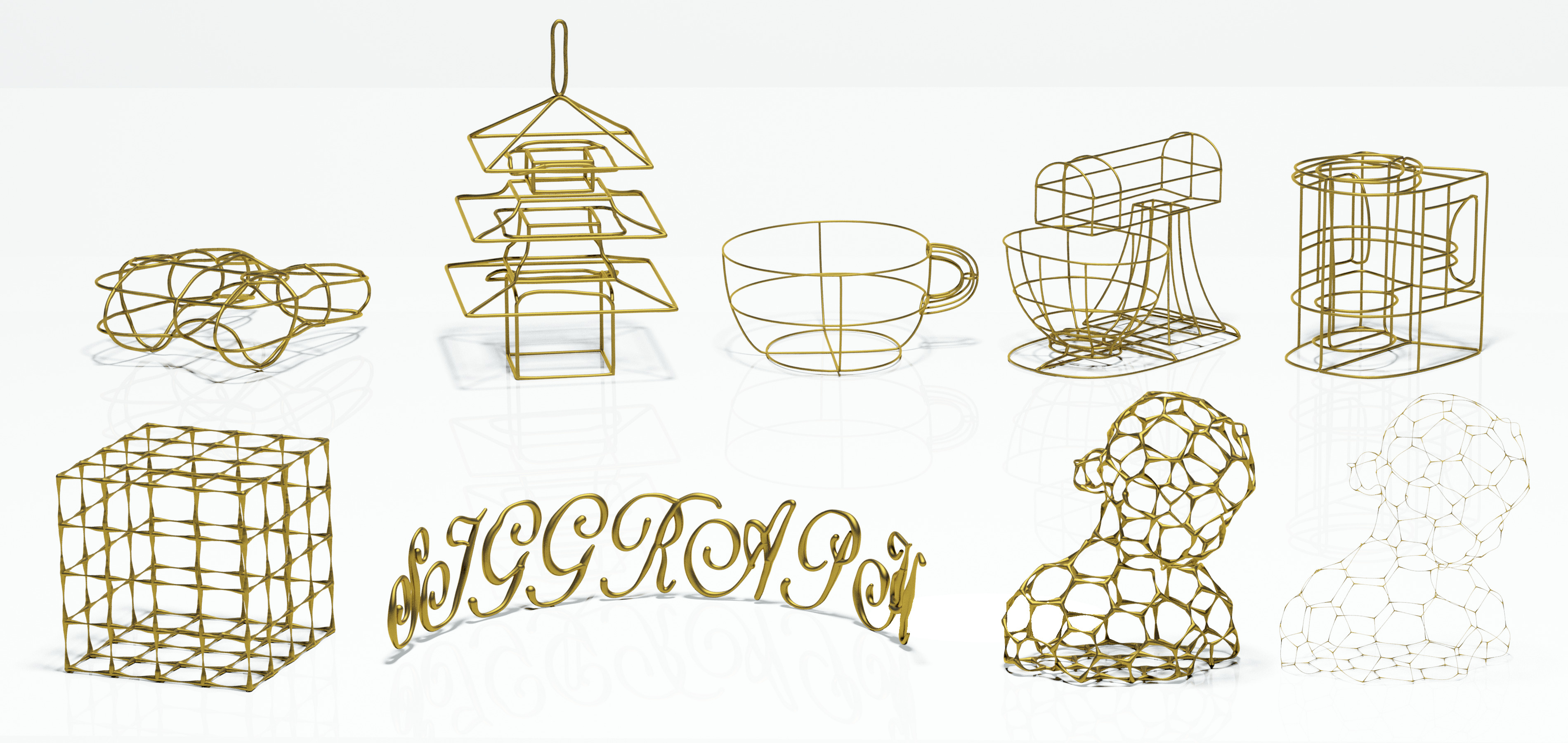}
     \caption{\tb{Reconstruction of synthetic models.} The synthetic dataset contains nine wire models. Four of these models, shown in the bottom row, have a varying thickness and are used for validation as summarized in~\tblref{evaluate_synthetic} and~\tblref{evaluate_synthetic_camErr}.}
\label{fig:synthetic_dataset}
\end{figure}

\subsection{Quantitative Evaluations}
\label{sec:metric_compare}
To quantitatively evaluate our method, we prepared a dataset of synthetic wire models, among which Cube, Text, Bimbo and Bimbo\_Thin models were created by ourselves and the others are from Mehra~\etal~\shortcite{mehra2009abstraction} and Xu~\etal~\shortcite{xu2014true2form}.
Four wire models (Cube, Text, Bimbo and Bimbo\_Thin) have varying thickness while the others have uniform thickness.
\figref{synthetic_dataset} shows the reconstruction of these digital models from the rendered images of the objects using a virtual camera moving along a simulated motion path. Only the intrinsic camera parameters are used as input; the camera path is estimated by our method and compared against the ground truth.

The following metrics are used for quality assessment:
%
\begin{itemize}
    \item {\em Reconstruction error (\RE)} measures the average of normalized closest point distances between the reconstructed model and the ground-truth in 3D.
 	\item {\em Relative reconstruction error (\RRE)} measures the average of normalized closest point distances between the reconstructed model and the ground-truth relative to the corresponding tube diameter of the closest point.
 	
    \item {\em Relative pose error (\RPE)} measures the pose error of an estimated camera path against its ground truth, following \cite{sturm2012benchmark}.
    
    \item {\em Projection error (\PE)} measures the average of normalized distances between the projection of points on the central curve and the closest points sampled on 2D central curves over all the frames.
    
    \item {\em Topology error} includes two metrics: {\em precision (\TPE)} and {\em recall (\TRE)}. \TPE \ measures the fraction of correctly reconstructed junctions in the reconstructed model and \TRE \ measures the fraction of correctly reconstructed junctions in the ground-truth model.
\end{itemize}
%
Note that the above distance measures in 3D and 2D are normalized by the diagonal length of the bounding box of the synthetic model in 3D and in the 2D projected view, respectively.

\begin{table}[!t]\tiny
    \centering
    \caption{Quantitative evaluation on reconstruction error (\RE); relative reconstruction error (\RRE); projection error (\PE); and topology error, including precision (\TPE) and recall (\TRE).}
    \resizebox{1.0\linewidth}{!}{
        \begin{tabular}{ccccccc}
            \toprule
            && {\RE} & {\RRE} & {\PE} & {\TPE} & {\TRE} \\
            \midrule
            Blender && 0.000596 & 0.0775 & 0.0009 & 86/87 & 86/86 \\
            Cup && 0.000530 & 0.0689 & 0.0008 &  25/26 & 25/25\\
            Game Pad && 0.000450 & 0.0505 & 0.0015 & 43/43 & 43/43 \\
            Japan House && 0.000557 & 0.0690 & 0.0024 &  48/49 & 48/50 \\
            Pencil Holder && 0.000578 & 0.0831 & 0.0011 &  45/46 & 45/45 \\
            Cube && 0.000511 & 0.0835 & 0.0010 & 98/98 & 98/98 \\
            Text && 0.000871 & 0.0960 & 0.0012 & 13/13 & 13/13 \\
            Bimbo && 0.000539 & 0.0636 & 0.0009 & 172/172 & 172/175 \\
            Bimbo\_Thin && 0.000298 & 0.1726 & 0.0008 & 174/174 & 174/175 \\
            \bottomrule
        \end{tabular}
    }
    \label{tbl:evaluate_synthetic}
\end{table}

\begin{table}[!t]\tiny
    \centering
    \caption{Quantitative evaluation on estimating camera pose error (\RPE).}
    \resizebox{0.85\linewidth}{!}{
        \begin{tabular}{cccc}
            \toprule
            && RPE ($\Delta=30$) &  Path length ($\Delta=30$) \\
            \midrule
            Blender && 0.0078 & 2.4848 \\
            Cup && 0.0031 & 1.1782 \\
            Game Pad && 0.0125 & 1.7534\\
            Japan House && 0.0401 & 2.9084 \\
            Pencil Holder && 0.0083 & 2.7369 \\
            Cube && 0.0109 & 1.3501 \\
            Text && 0.0119 & 0.6947 \\
            Bimbo && 0.0066 & 1.9390 \\
            Bimbo$\_$Thin && 0.0031 & 1.9453 \\
            \bottomrule
        \end{tabular}
    }
    \label{tbl:evaluate_synthetic_camErr}
\end{table}

\tblref{evaluate_synthetic} reports the statistics on the reconstruction errors. It shows that our algorithm produces faithful 3D reconstruction with correct recovery of 3D topology. 
The reconstruction errors (\RE) are less than $0.001$ and the projection errors (\PE) are less than $0.003$. The relative reconstruction errors (\RRE) are less than 10\% except for the Bimbo\_Thin model, whose curve width, after projecting to the images, is 1-3 pixels, while the width of the other models ranges from 5 to 10 pixels.
%
Table~\ref{tbl:evaluate_synthetic_camErr} reports the relative pose errors (RPE)~\cite{sturm2012benchmark} measured on the nine estimated camera paths. Note that the RPE is defined with respect to the frame interval $\Delta$, which is set to 30 in this evaluation. That is, the RPE here is the averaged camera pose error (i.e. drifting) during the span of 30 frames along the camera motion path.
Table~\ref{tbl:evaluate_synthetic_camErr} also reports the averaged length of camera displacement during the span of 30 frames. It follows that, on average, the relative camera pose error accumulated over 30 frames is less than $2\%$ of the average length of camera motion during 30 frames. 

%


\begin{figure}[!t]
	\mpage{0.31}{\includegraphics[width=\linewidth]	{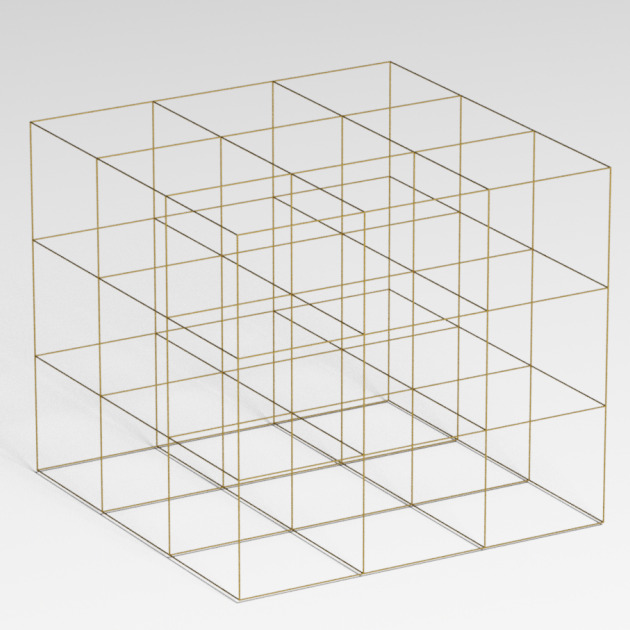}}
	\mpage{0.31}{\includegraphics[width=\linewidth]	{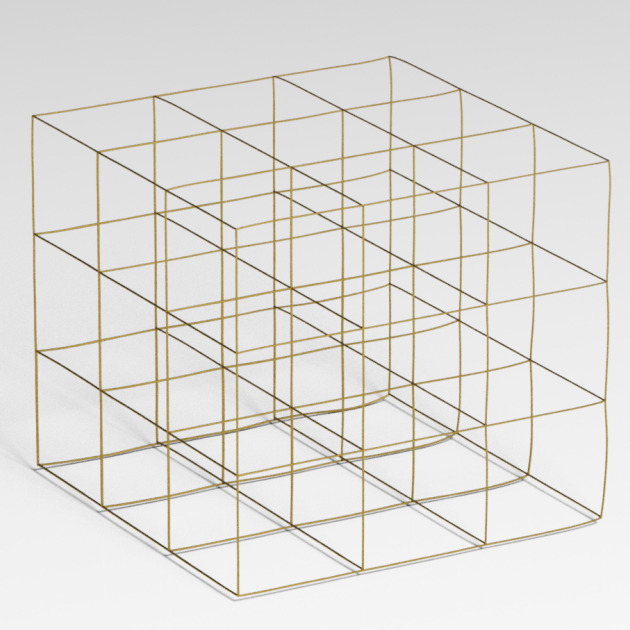}}
    \mpage{0.31}{\includegraphics[width=\linewidth]	{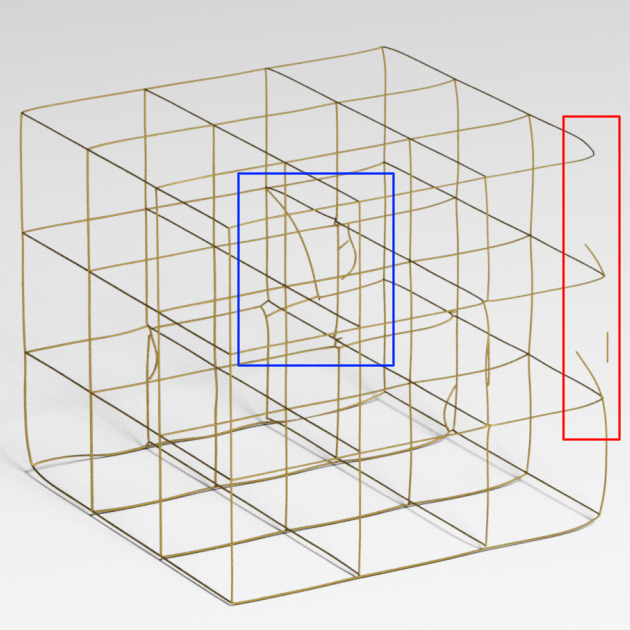}}
     \\
    \mpage{0.31}{(a)}
    \mpage{0.31}{(b)}
    \mpage{0.31}{(c)}
    
    \caption{\textbf{Effect of curve matching.} (a) Input synthetic $3\times 3\times 3$ grid model. (b) Reconstruction using our curve matching algorithm. (c) Reconstruction using a naive closest point search algorithm, yielding notable artifacts, such as missing parts (red box) and redundant parts (blue box).
    }
    \label{fig:compare_curve_matching}
\end{figure}

\subsection{Ablation Studies}

\paragraph{Effect of Curve Matching}
The tailored curve matching algorithm discussed in~\secref{iteration} plays a crucial part in reconstructing accurate curve skeletons.
To validate its effectiveness, we conducted an experiment that compares our curve matching method with a na\"ive closest point search algorithm on reconstructing a synthetic $3 \times 3 \times 3$ grid model (see~\figref{compare_curve_matching}(a)).
As shown in~\figref{compare_curve_matching}(c), the na\"ive method caused obvious artifacts such as missing and redundant parts, while our method produces high-fidelity reconstruction (see~\figref{compare_curve_matching}(b)).


\begin{figure}[!b]
	\mpage{0.31}{\includegraphics[width=\linewidth]	{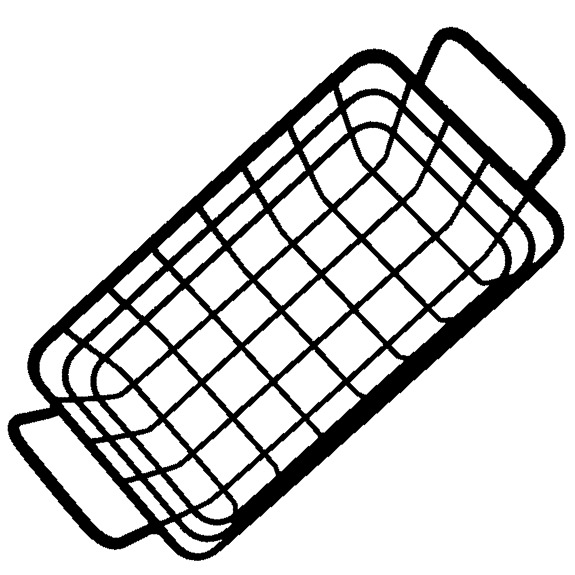}}
    \mpage{0.31}{\includegraphics[width=\linewidth]{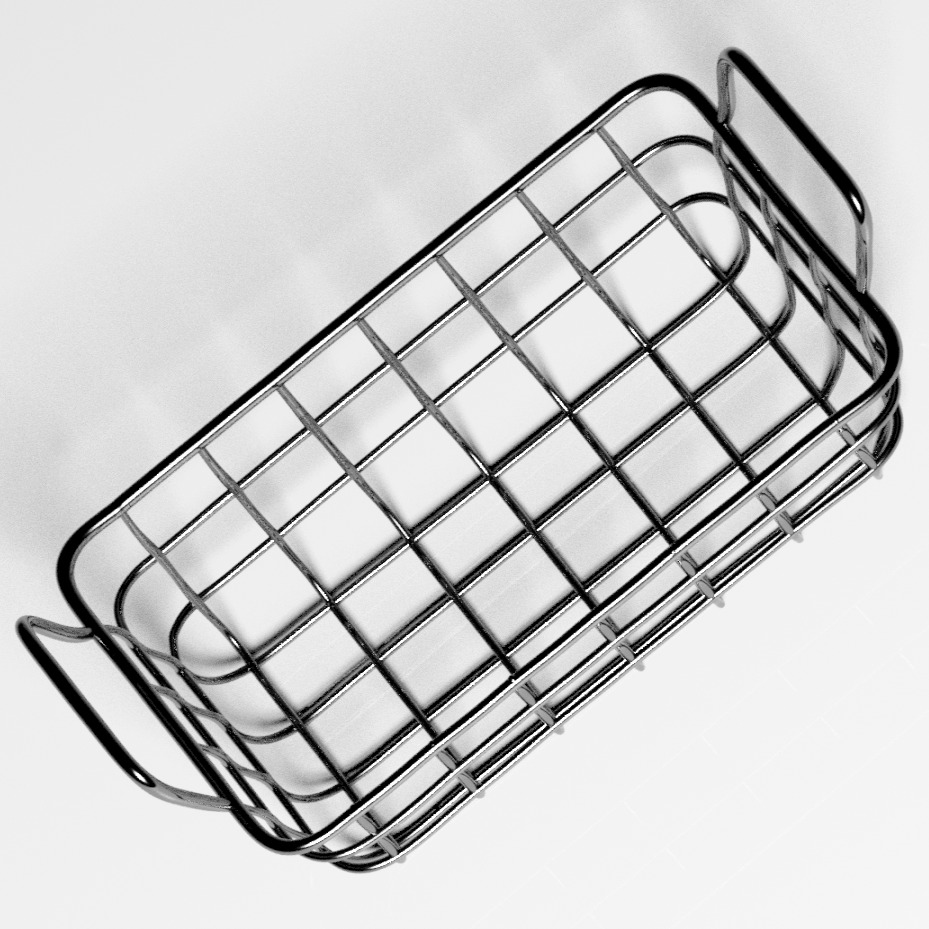}}
    \mpage{0.31}{\includegraphics[width=\linewidth]{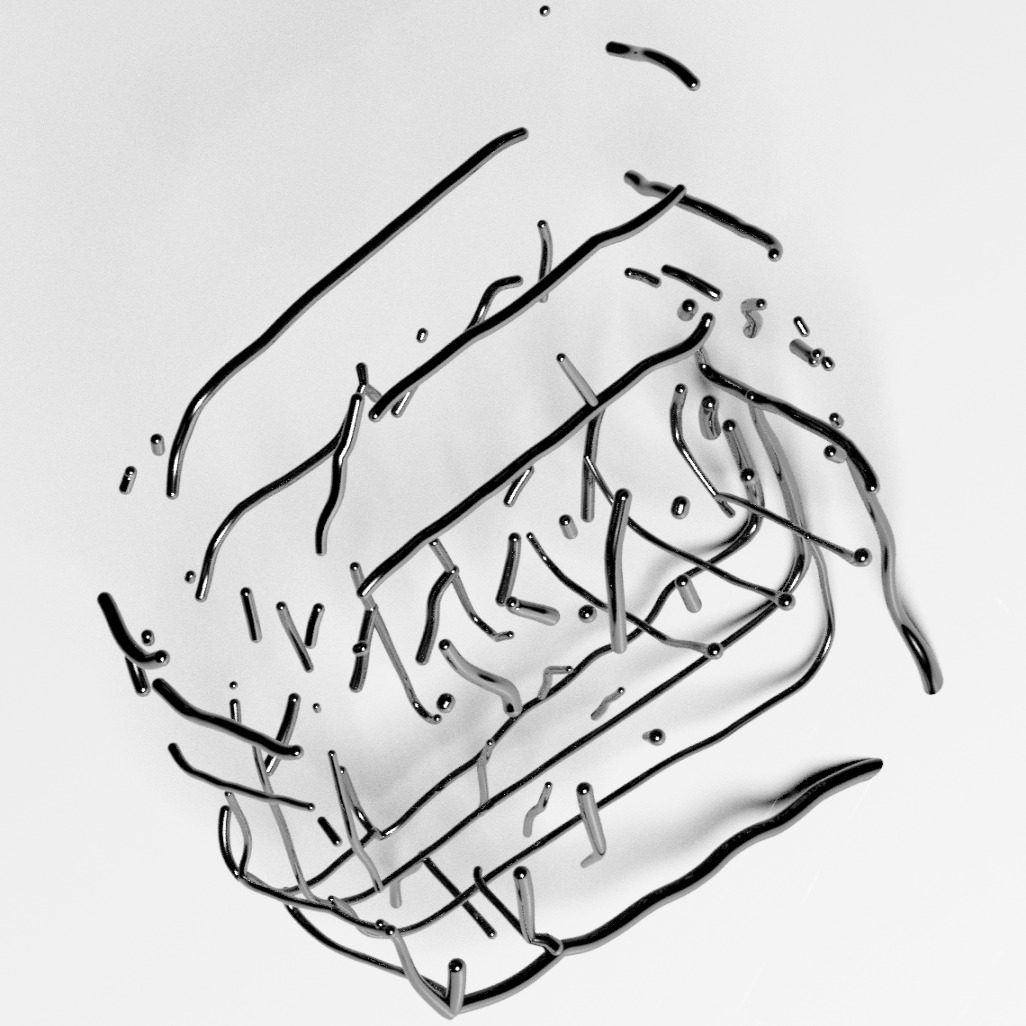}}
    \\
    \mpage{0.31}{(a)}
    \mpage{0.31}{(b)}
    \mpage{0.31}{(c)}
    
    \caption{\textbf{Effect of self-occlusion detection.} (a) An input frame with severe self-occlusion. (b) Reconstruction result with self-occlusion detection. (c) Reconstruction result without self-occlusion detection.}
    
    \label{fig:compare_occlusion}
\end{figure}

\paragraph{Effect of Self-occlusion Detection}
As discussed in~\secref{occlusion}, the ability to detect and handle self-occlusion is key to robust reconstruction of thin structures.
To show the impact of self-occlusion detection on the final reconstruction, we took the video of a real basket model. This input video has severe self-occlusion in over half of its frames (see~\figref{compare_occlusion}(a)).
We tested our system with this input video with and without enabling the self-occlusion detection, and show 
the respective reconstruction results in~\figref{compare_occlusion}(b) and ~\figref{compare_occlusion}(c). Clearly, there are significant reconstruction errors in both geometry and topology when self-occlusion detection is turned off. 
%



\subsection{Sensitivity to Camera Shake}
\begin{figure}[!t]
    
  \mpage{0.31}{\includegraphics[width=\linewidth]{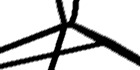}}
  %
  \mpage{0.31}{\includegraphics[width=\linewidth]{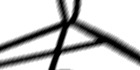}}
  %
  \mpage{0.31}{\includegraphics[width=\linewidth]{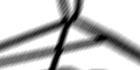}}
  \\
  \mpage{0.31}{\includegraphics[width=\linewidth]{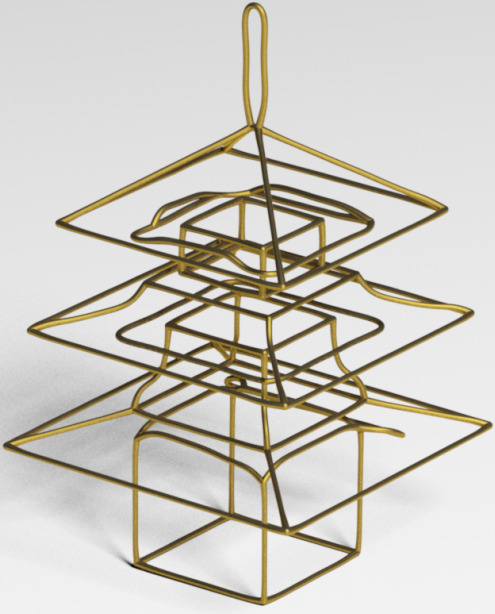}}
    \mpage{0.31}{\includegraphics[width=\linewidth]{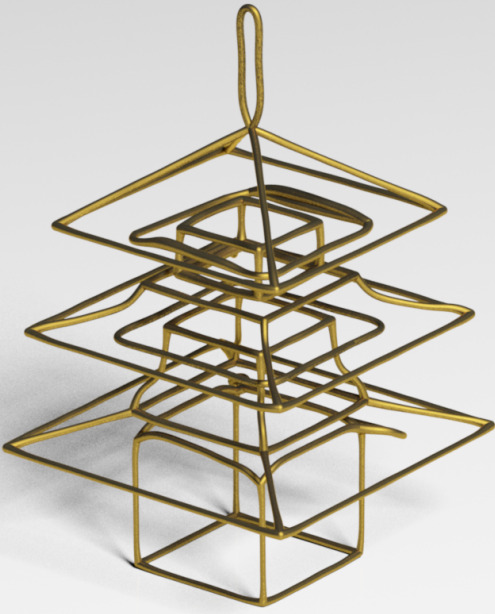}}
    \mpage{0.31}{\includegraphics[width=\linewidth]{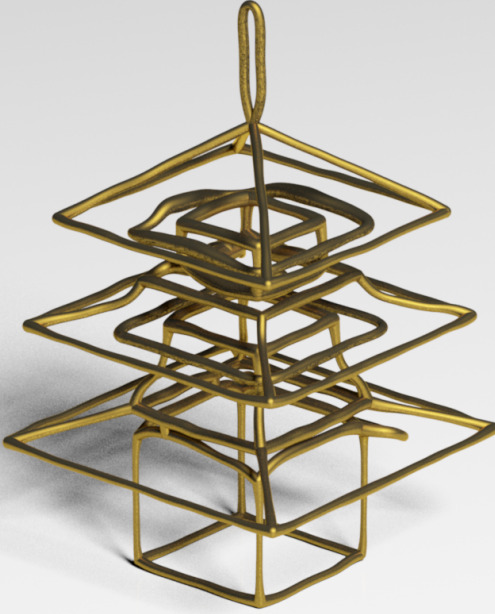}}
    \\
    \mpage{0.31}{\small Cam. shake (small)}
 \mpage{0.31}{\small Cam. shake (medium)}
 \mpage{0.31}{\small Cam. shake (large)}
 \caption{\tb{Sensitivity to camera shake.} (Top) Blurry input frames due to different degrees of camera shake. (Bottom) Corresponding reconstruction results.}
    \label{fig:camera_shake_result}
\end{figure}

Camera shake caused by unsteady hands or other motions during video capturing leads to blurry input images.
To quantitatively evaluate how well our method could resist camera shake, we conducted an experiment using synthetic 3D wire models. Specifically, we rendered 200 images of the Japan House model and used a non-uniform blurring model~\cite{whyte:ijcv:2012} to simulate camera shake effect as shown in~\figref{camera_shake_result}(top).
With this blurring model, the rotation of a camera is represented as Euler angle and we added different degrees of perturbations to each angle by sampling a Gaussian distribution with zero mean and varying standard deviations of 0.1$^{\circ}$ (small), 0.3$^{\circ}$ (medium) and 0.5$^{\circ}$ (large).
\figref{camera_shake_result} shows the reconstruction results with different degrees of image blurring and the quantitative results are reported in~\tblref{evaluate_synthetic_segmentation},
We can see that our method can resist a certain degree of camera shake.
In the case of large camera shake, the blurry boundaries cause relatively large estimation errors in image segmentation, which in turn results in increased reconstruction errors in terms of both RE and RRE.

\begin{figure}[!b]

  \mpage{0.31}{\includegraphics[width=\linewidth]{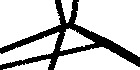}}
    \mpage{0.31}{\includegraphics[width=\linewidth]{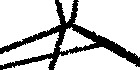}}
    \mpage{0.31}{\includegraphics[width=\linewidth]{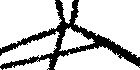}}
    \\
  \mpage{0.31}{\includegraphics[width=\linewidth]{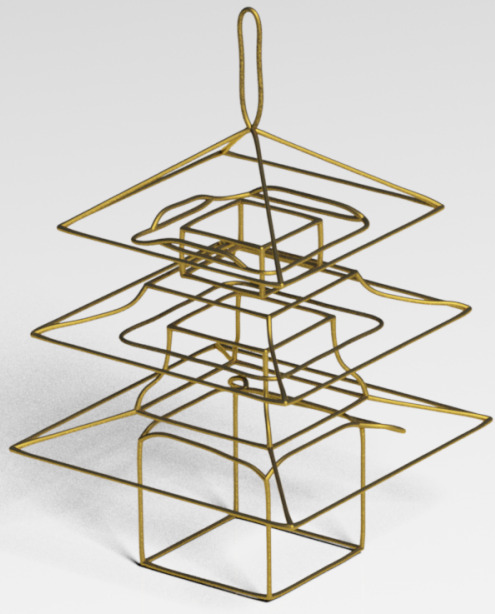}}
    \mpage{0.31}{\includegraphics[width=\linewidth]{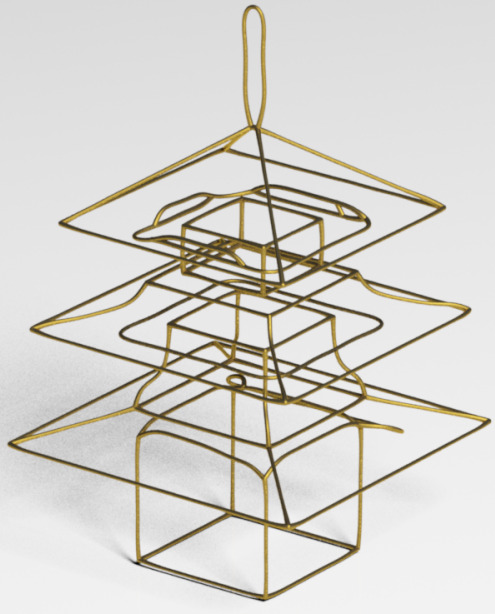}}
    \mpage{0.31}{\includegraphics[width=\linewidth]{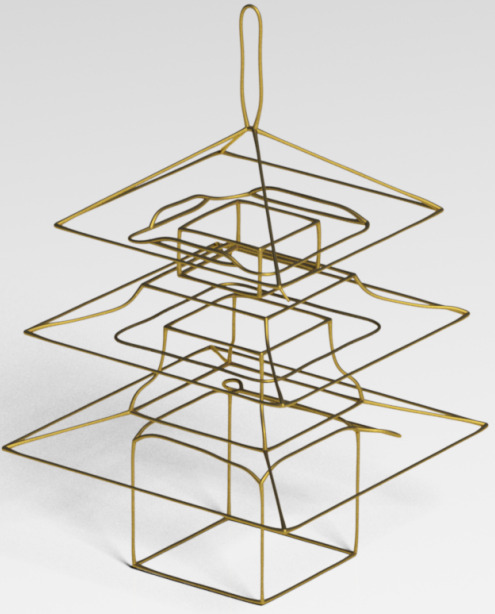}}
    \\
    \mpage{0.31}{Seg. noise (low)}
 \mpage{0.31}{Seg. noise (medium)}
  \mpage{0.31}{Seg. noise (high)}
    \caption{\tb{Sensitivity to segmentation noise.} (Top) The input segmentation masks after applying different among of boundary noise. (Bottom) Our reconstruction results.}
    \label{fig:segmentation_result}
\end{figure}

\begin{table}[!t]\tiny
    \centering
    \caption{Quantitative evaluation on the sensitivity to different degrees of camera shake (upper block) and segmentation noises (lower block). Please refer to~\secref{metric_compare} for the definition of error metrics.}
    \resizebox{1.0\linewidth}{!}{
        \begin{tabular}{ccccccc}
            \toprule
            && RE & RRE & PE & TPE & TRE \\
            \midrule
            Cam. shake (small) && 0.000651 & 0.0807 & 0.0014 & 48/48 & 48/50 \\
            Cam. shake (medium) && 0.001144 & 0.1418 & 0.0024 & 44/47 & 44/50\\
            Cam. shake (large) && 0.003349 & 0.4163 & 0.0043 & 41/47 & 41/50\\
            \noalign{\vskip 1pt}
            \hline
            \noalign{\vskip 2pt}
            Seg. noise (low) && 0.000765 & 0.0947 & 0.0009 & 48/49 & 48/50 \\
            Seg. noise (medium) && 0.000891 & 0.1102 & 0.0009 & 50/52 & 50/50 \\
            Seg. noise (high) && 0.001168 & 0.1445 & 0.0008 & 50/51 & 50/50 \\
            \bottomrule 
        \end{tabular}
    }
    \label{tbl:evaluate_synthetic_segmentation} 
\end{table}

\begin{figure*}[!t]
\mpage{0.19}{\includegraphics[width=\linewidth]{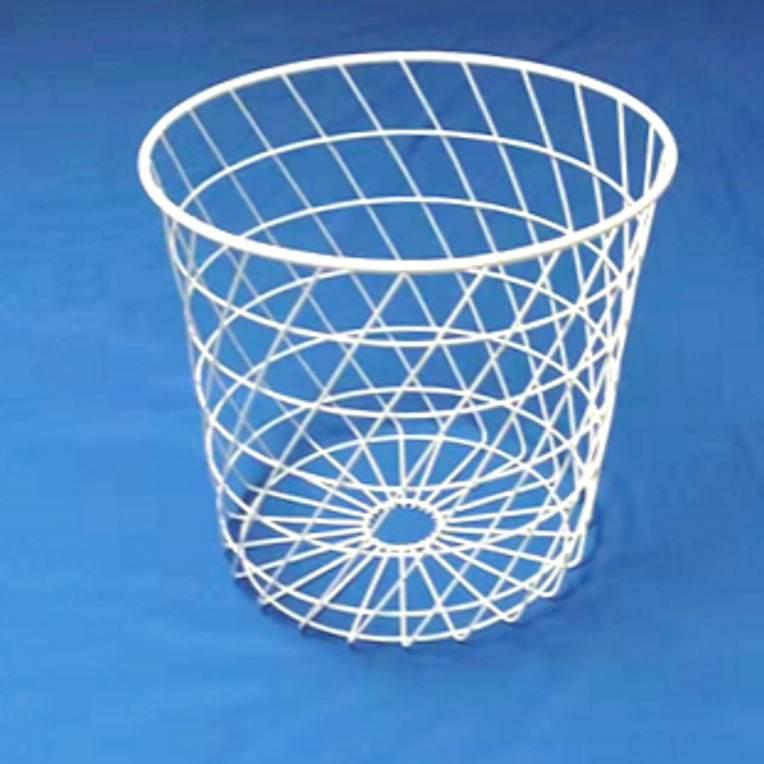}} 
\mpage{0.19}{\includegraphics[width=\linewidth]{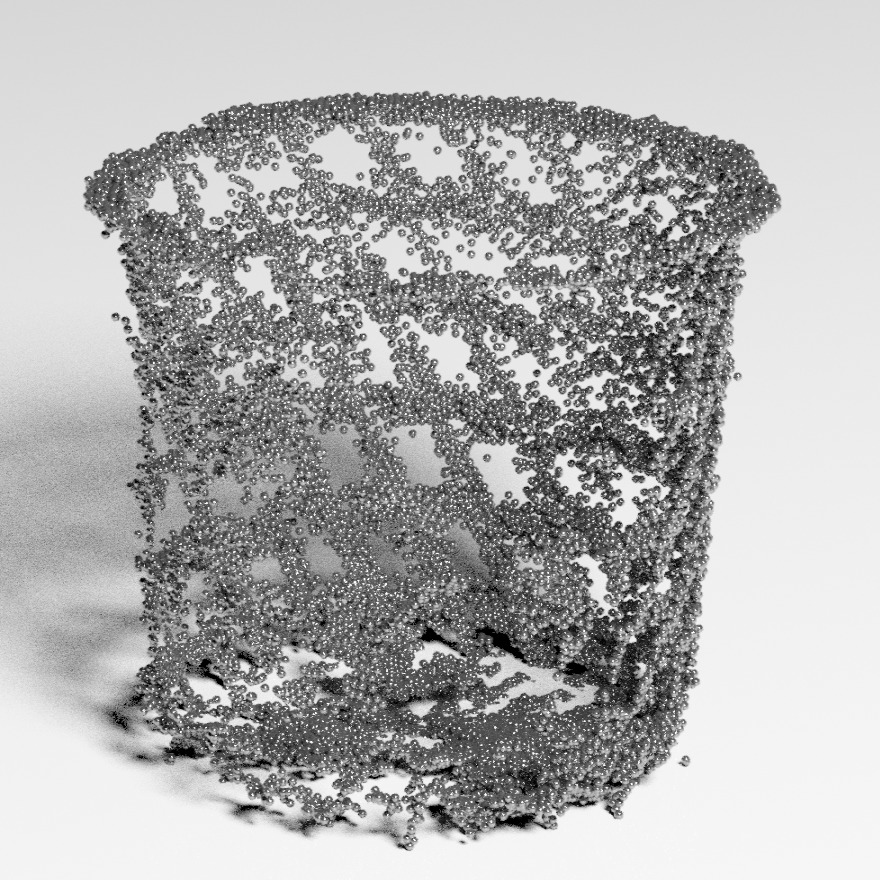}}
\mpage{0.19}{\includegraphics[width=\linewidth]{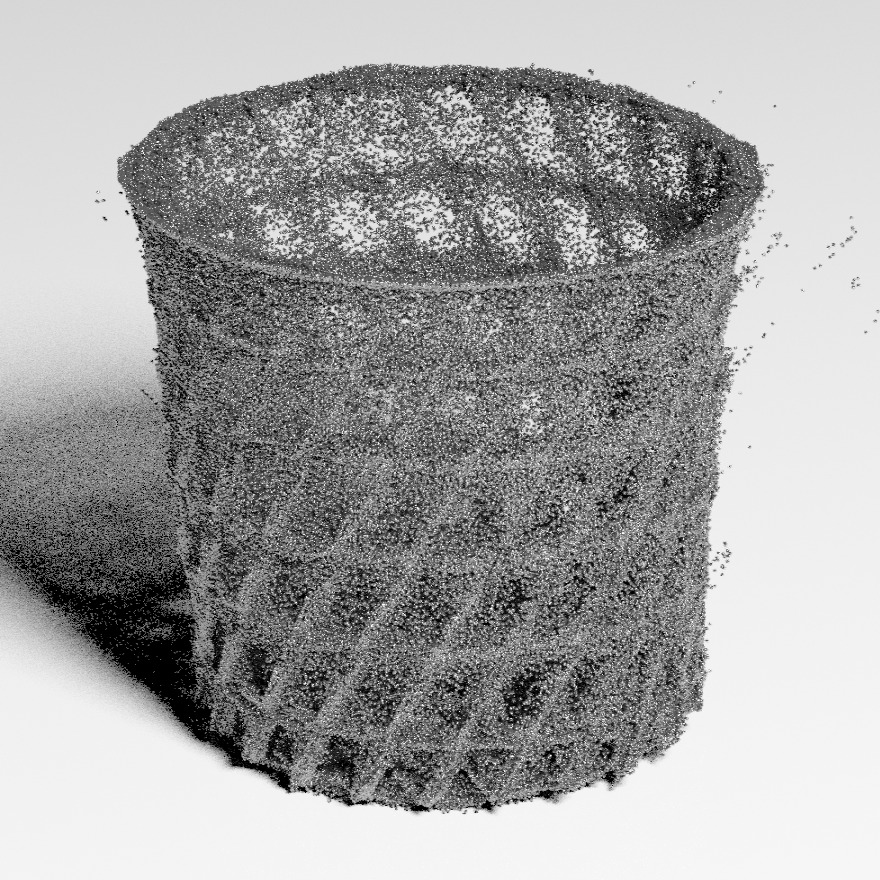}}
\mpage{0.19}{\includegraphics[width=\linewidth]{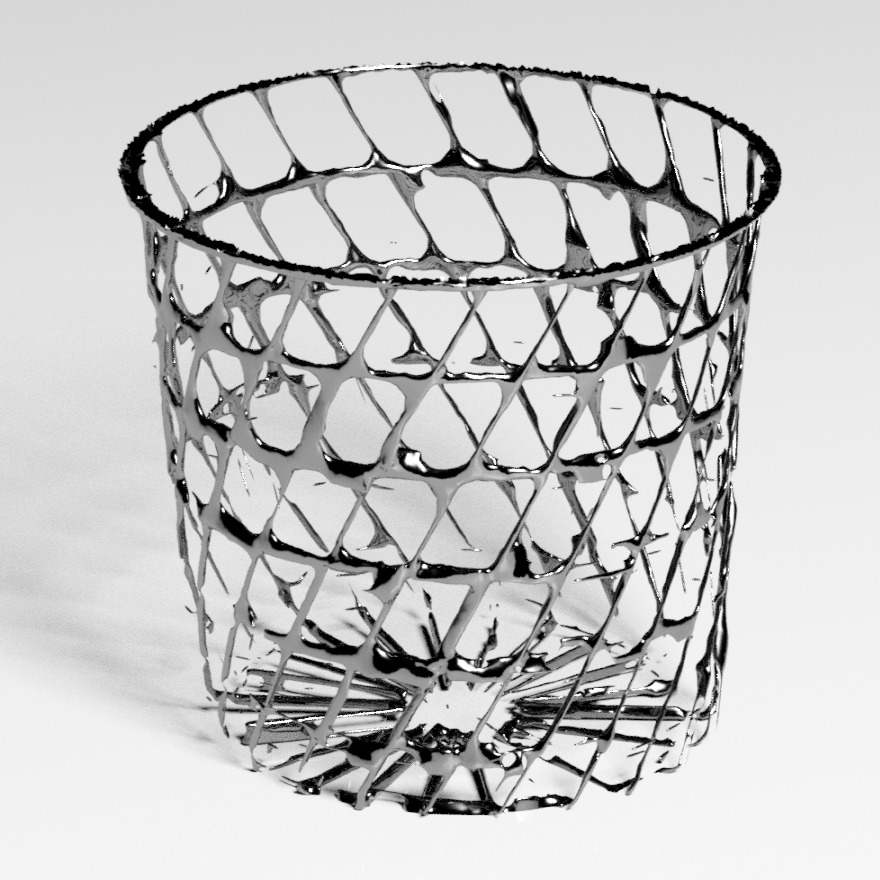}}
\mpage{0.19}{\includegraphics[width=\linewidth]{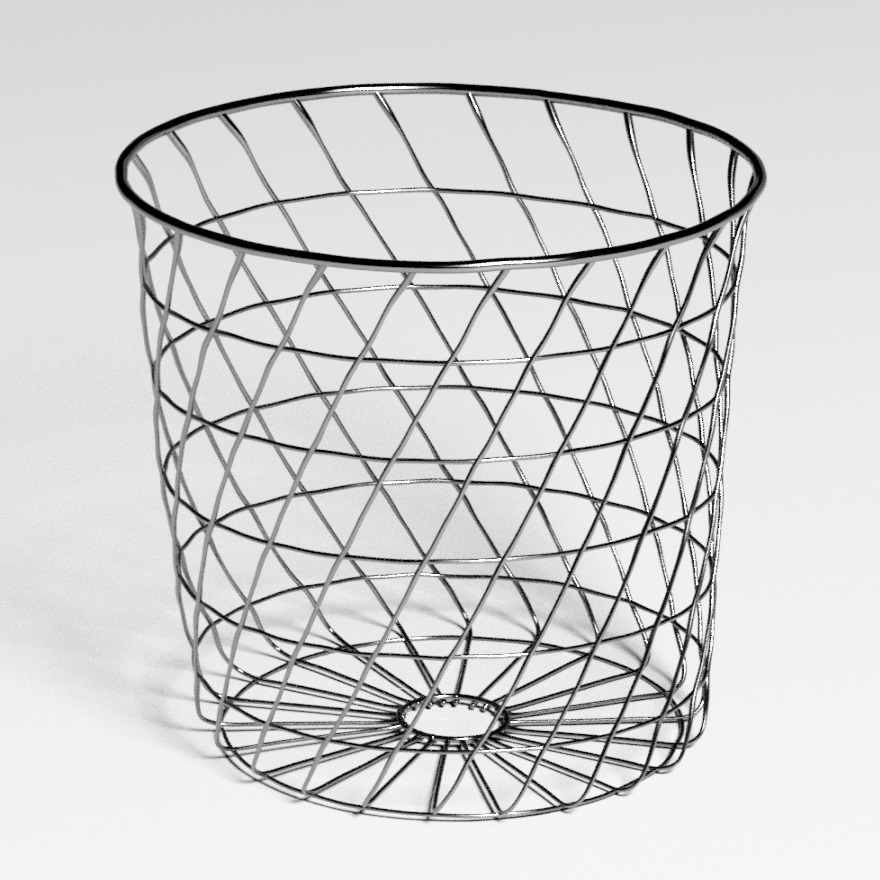}}
\\
\mpage{0.19}{\includegraphics[width=\linewidth]{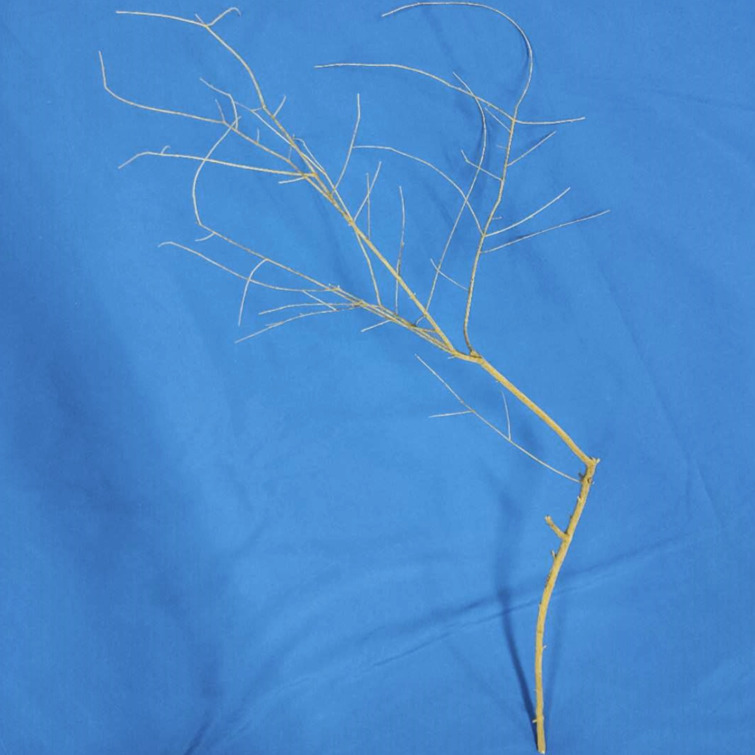}}
\mpage{0.19}{\includegraphics[width=\linewidth]{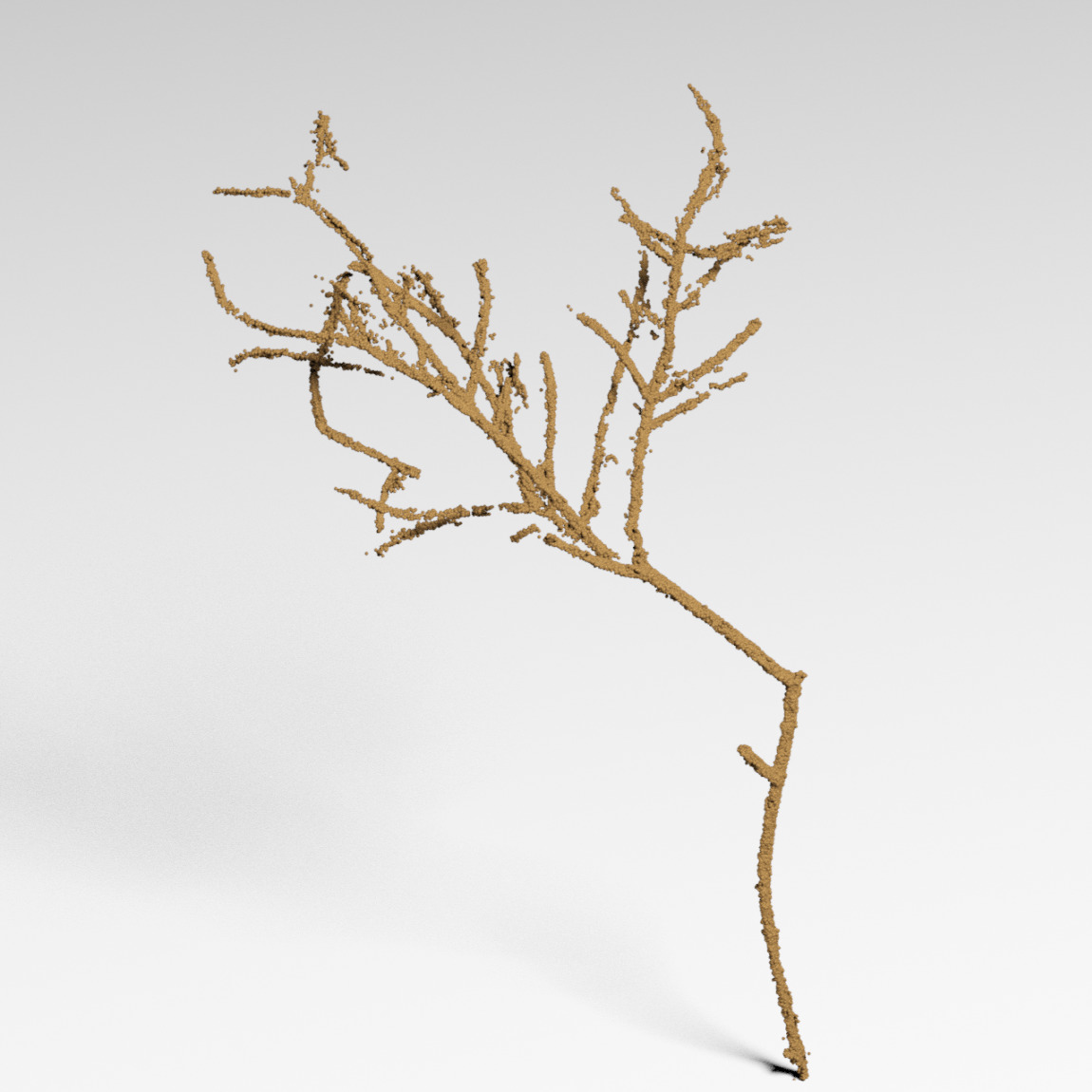}}
\mpage{0.19}{\includegraphics[width=\linewidth]{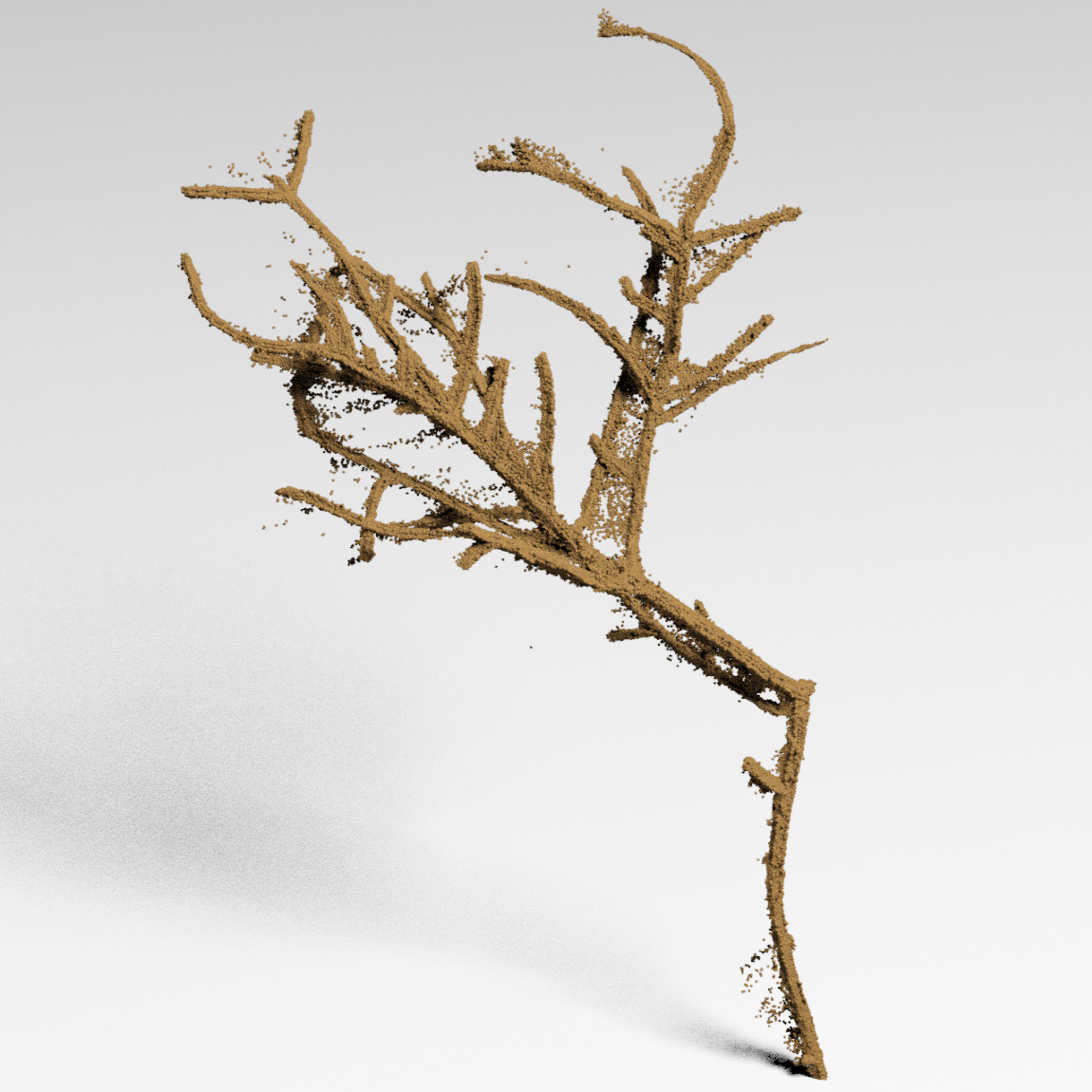}}
\mpage{0.19}{\includegraphics[width=\linewidth]{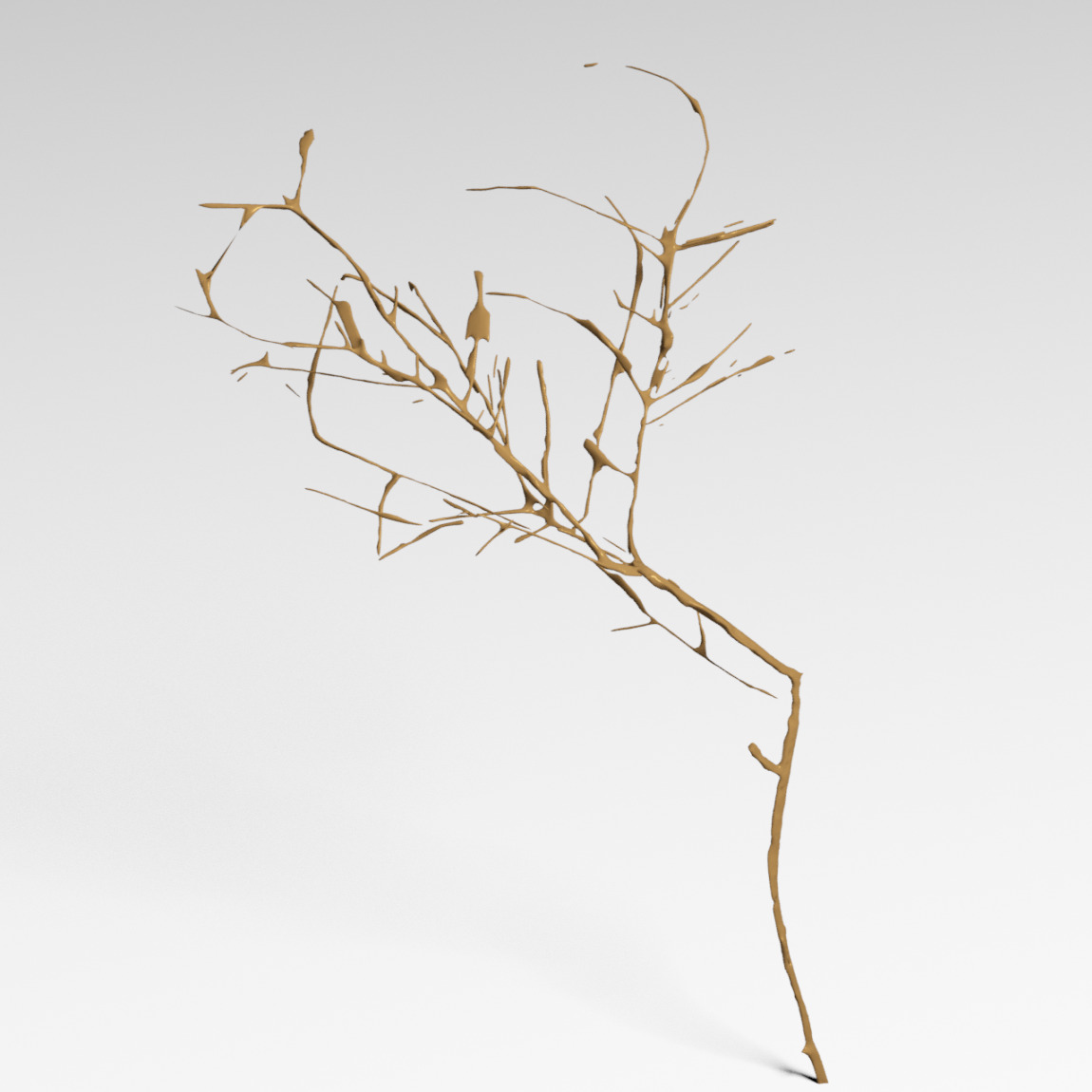}}
\mpage{0.19}{\includegraphics[width=\linewidth]{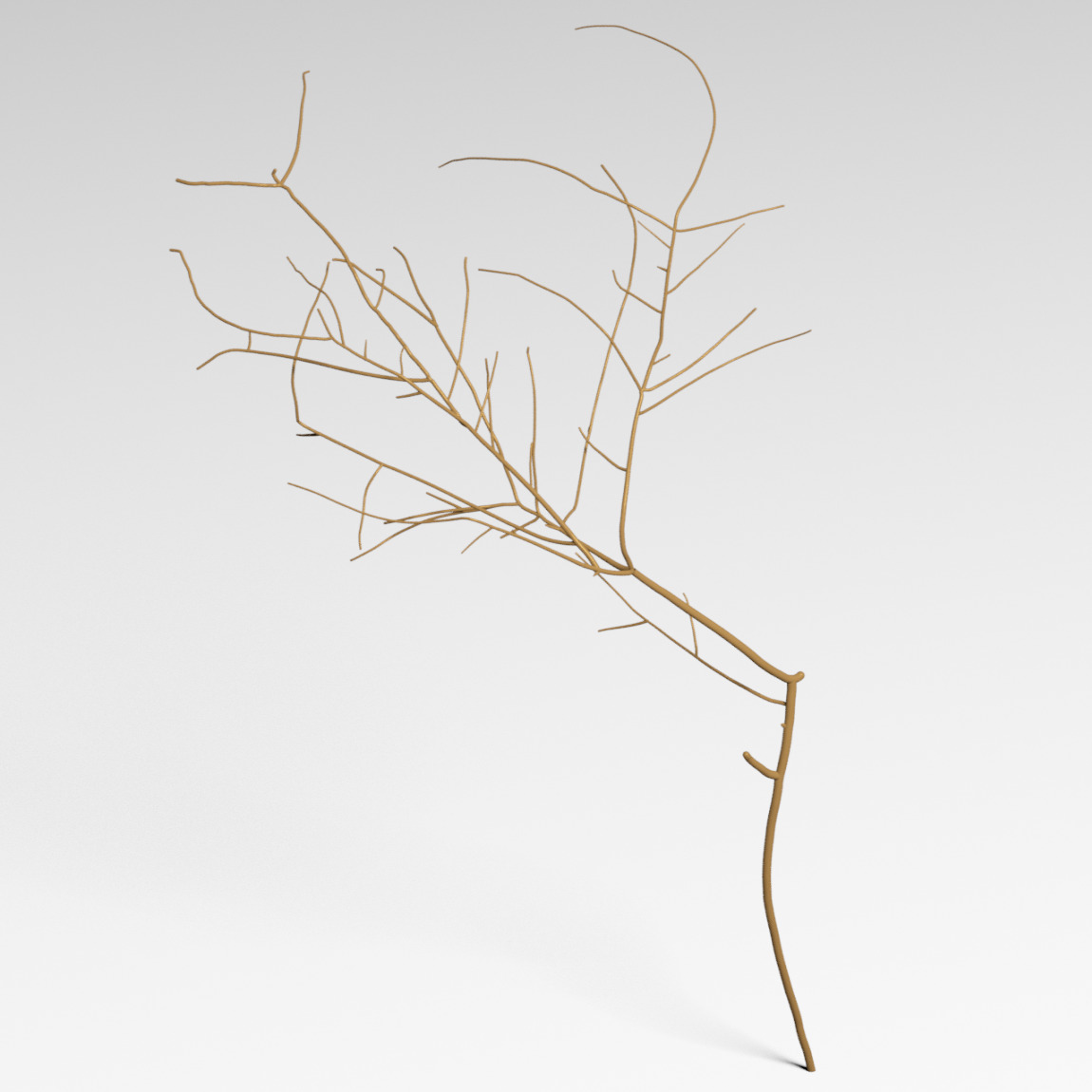}}
\\
\mpage{0.19}{Reference Image}
\mpage{0.19}{PMVS}
\mpage{0.19}{COLMAP}
\mpage{0.19}{\citet{li:cvpr:2018}}
\mpage{0.19}{Ours}
\\
     \caption{Comparisons with three multi-view stereo (MVS) algorithms: PMVS~\cite{Furukawa:pami:2010},  COLMAP~\cite{schonberger:cvpr:2016,schonberger:eccv:2016}, and Li~\etal~\cite{li:cvpr:2018}.}
\label{fig:compare_large}
\end{figure*}

\subsection{Sensitivity to Segmentation Noise}
%
To evaluate the sensitivity of our method to the quality of the segmentation mask, we conducted an experiment that adds different levels of noise to the segmentation boundary of the Japan House model (see~\figref{segmentation_result}(top)). 
The noise is added by sampling a Gaussian distribution with zero mean and standard deviations of 0.3 (low), 0.5 (medium) and 0.7 (high).
The reconstruction results are shown in~\figref{segmentation_result}(bottom) and the quantitative results are reported in~\tblref{evaluate_synthetic_segmentation}.
The results demonstrate the robustness of our method to a certain level of segmentation noise.

\subsection{Comparisons with Baselines}
\label{sec:baseline_compare}
We first compare our method with three multi-view stereo (MVS) algorithms, PMVS~\cite{Furukawa:pami:2010}, COLMAP~\cite{schonberger:cvpr:2016,schonberger:eccv:2016}, and Li~\etal~\cite{li:cvpr:2018}, on a bucket model and a model of leafless tree branches
%
The number of input frames used as input for all these methods is 182 for each object.
%
Note that these methods all require rich background texture features for camera pose estimation. Therefore, for fair comparison, their input videos are taken to contain some textured objects in the background.
As shown in~\figref{compare_large}, the reconstructions by these methods contain significant noise and missing parts.
%
In addition to visual comparison, we also conducted a quantitative evaluation on the bucket model shown in~\figref{compare_large} (top row).
For the quality assessment, we employ the commonly used re-projection error that measures the average distance between the projected 3D reconstruction and 2D image segmentation over all the input frames.
The error of our reconstruction is less than 0.0007, which is smaller than the errors of the other three methods:  Li~\etal~(0.0015), PMVS (0.0034), and COLMAP (0.0023). Here the error value is normalized by the diagonal length of the 2D bounding box of the wire model in each 2D projected view.



\begin{figure}[!b]
\mpage{0.3}{\includegraphics[width=\linewidth]	{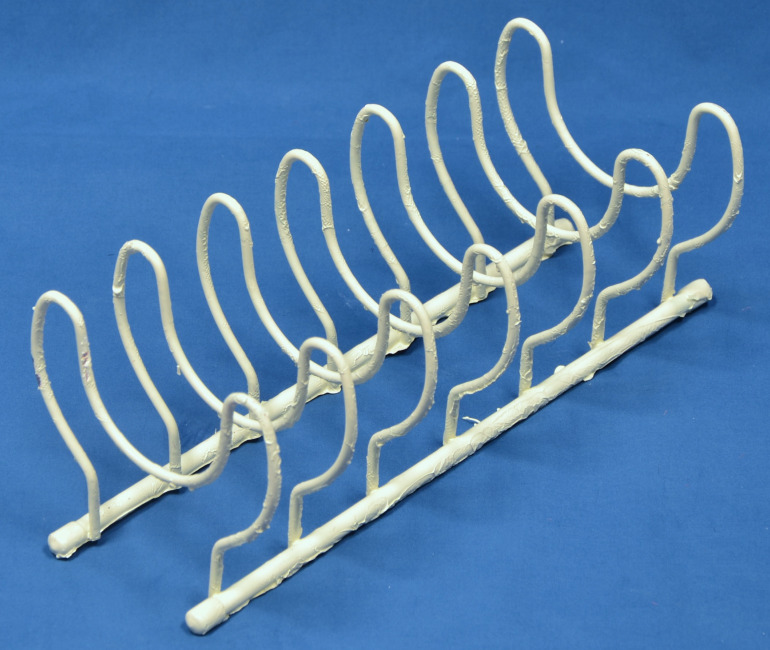}}
\mpage{0.3}{\includegraphics[width=\linewidth]	{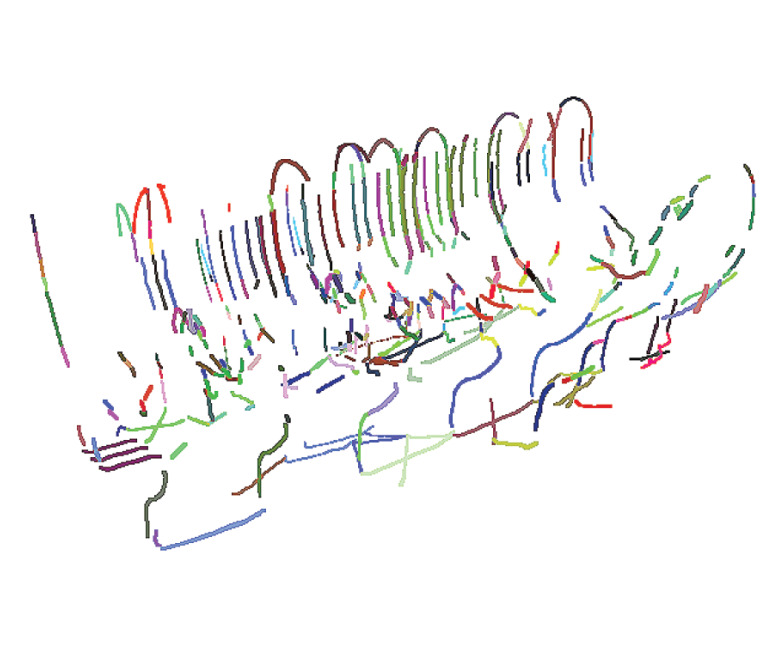}}
\mpage{0.3}{\includegraphics[width=\linewidth]	{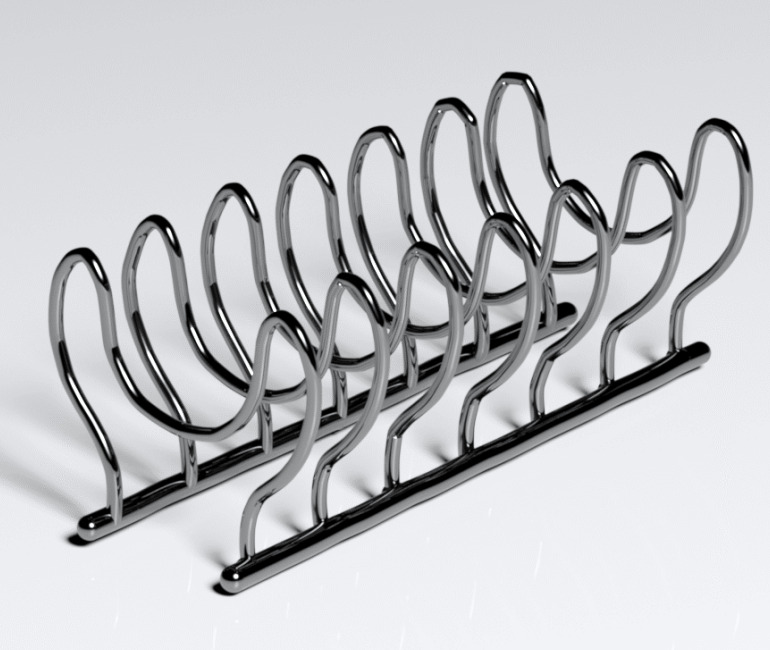}}
\\
\mpage{0.3}{Reference Image}
\mpage{0.3}{\citet{Liu:sigg:2017}}
\mpage{0.3}{Ours}
\\
     \caption{Comparison with~\citet{Liu:sigg:2017}}
\label{fig:compare_wireart}
\end{figure}



We further compare our method with a recent image-based method for 3D wire reconstruction~\cite{Liu:sigg:2017}, which uses three images along with intrinsic camera parameters and camera poses as input.
In our experiment the three input images were captured using the same viewpoint setting reported in their paper to avoid self-occlusion as much as possible in these views.

%
As shown in~\figref{compare_wireart} (middle), the method in~\cite{Liu:sigg:2017} fails for a wire model with a repetitive curve structure, while our method reconstructs the model successfully. 

\begin{figure}[!t]
\mpage{0.45}{\includegraphics[width=\linewidth]	{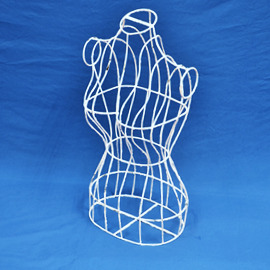}}
\mpage{0.45}{\includegraphics[width=\linewidth]	{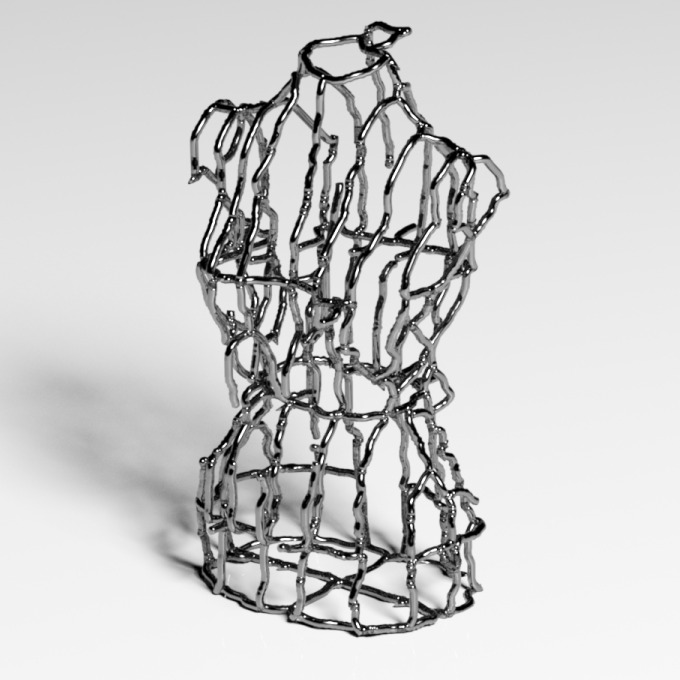}}
\\
\mpage{0.45}{Reference Image}
\mpage{0.45}{\citet{Tabb:cvpr:2013}}
\\
\mpage{0.45}{\includegraphics[width=\linewidth]	{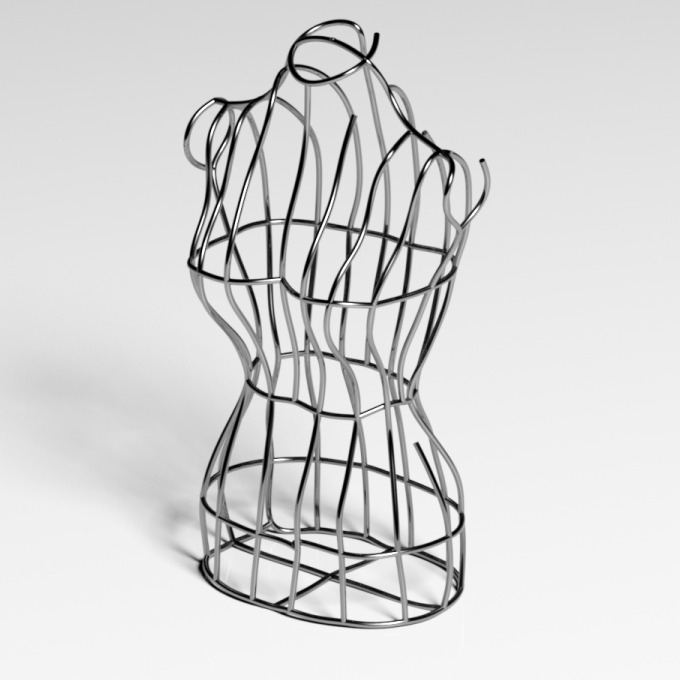}}
\mpage{0.45}{\includegraphics[width=\linewidth]	{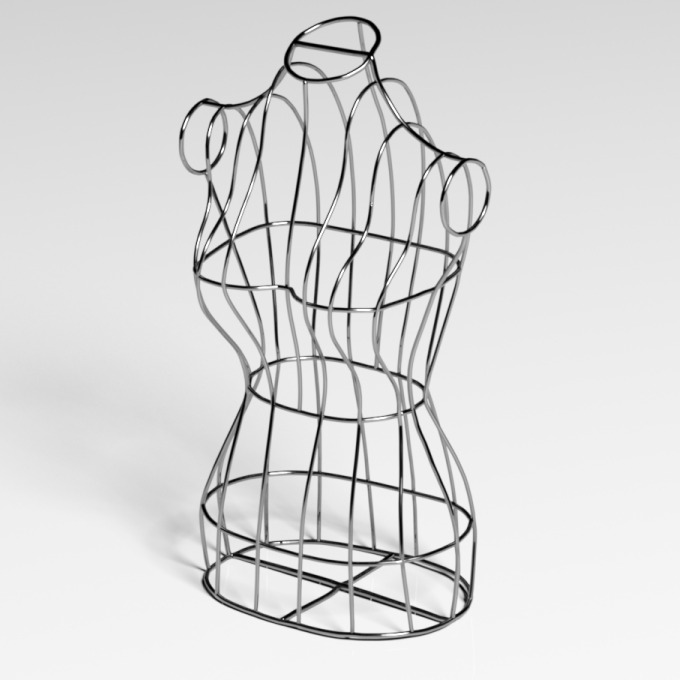}}
\\
\mpage{0.45}{CurveFusion}
\mpage{0.45}{Ours}

\caption{Comparison with Tabb~\citep{Tabb:cvpr:2013} and CurveFusion~\cite{Liu:sigga:2018}.}
\label{fig:compare_tabbcurvefusion}
\end{figure}

\figref{compare_tabbcurvefusion} shows a comparison to CurveFusion~\cite{Liu:sigga:2018}, an RGBD video-based method for thin structure reconstruction, and a state-of-the-art visual-silhouette based method by Tabb~\cite{Tabb:cvpr:2013}. %
%
For the CurveFusion method, the input is an RGBD sequence consisting of 261 frames. 
For the visual-silhouette based method, the author helped us to take photos of the wire model from 144 pre-calibrated camera positions.
It can be seen that the reconstruction by~\cite{Tabb:cvpr:2013} is noisy and irregular. Although the CurveFusion method shows impressive reconstruction quality, its result contains obvious defects, such as missing curves and incorrect topology, caused by the limited resolution of the commodity depth sensor used. 
Moreover, well-known limitations of the infrared-based depth sensors are that it does not work outdoor in strong sunlight and that objects with black surface cannot be scanned. 
In contrast, our method is able to handle these situations and produces reconstruction results with superior quality.

\subsection{Performance}
\label{sec:timing}
Here we report the performances of our method on a machine with Intel i5-8300H 2.3 GHz CPU and 16G RAM.
The overall computation time is proportional to the number of input images and the complexity of wire models.
For a simple wire model, such as the Fat Cat model (see~\figref{visual_results}(bottom-left)) with 124 input frames, our method took 138 seconds to reconstruct the model.
The computation times increases to 25 minutes for reconstructing a more complex example, such as the Bucket model (see~\figref{teaser}(left)) with 229 input frames.
For a breakdown, the initialization step (\secref{initialize}) and surface reconstruction step (\secref{surf_reconstruct}) take only 3\% and 2\% of the total time, respectively.
The main bottleneck lies in the process of iterative structure optimization  (\secref{iteration}), which takes about 95\% of the total running time.
%

\subsection{Limitations}
\label{sec:limits}
%
%
Our method requires that an input video be captured with a simple or clean background for easy and accurate foreground segmentation. Therefore, future work is needed to enable robust foreground segmentation of wire models against a general background. Another limitation is that our method assumes that wire models are made of tubular wires with circular sections. Therefore it cannot accurately reconstruct wire models with non-circular sections. 

\section{Conclusions}
We have proposed a new method for high quality reconstruction of 3D thin structures from an RGB video. Our method does not require camera poses as input but uses only curve feature correspondence to accurately estimate the camera poses of the input image frames. This allows us to reconstruct more complex thin structure in higher quality than the previous method. As future work, we would like to study the segmentation of thin structures against a general background so to be able to apply the proposed reconstruction pipeline in a natural setting. Another extension is to consider the reconstruction of general objects consisting of thin structure components as well as extended surfaces. 
\label{sec:conclusions}

\begin{acks}
We would like to thank the reviewers for their valuable and insightful comments. We also thank Shiwei Li, Amy Tabb, Rhaleb Zayer for their help with experiments. This work was partially funded by the Research Grant Council of Hong Kong (GRF 17210718), ERC Consolidator Grant 770784, Lise Meitner Postdoctoral Fellowship, Ministry of Science and Technology of Taiwan (108-2218-E-007-050 and 107-2221-E-007-088-MY3).
\end{acks}

\bibliographystyle{ACM-Reference-Format}
\bibliography{main}

\end{document}